\documentclass[12pt]{article}
\usepackage{amsfonts,amsthm,amsmath,amssymb,upgreek,bm}
\usepackage{dsfont}
\usepackage[paper=letterpaper,margin=.85in]{geometry}
\usepackage{graphicx}
\usepackage{units}
\usepackage{float}
\usepackage[export]{adjustbox}
\usepackage{bbold}
\usepackage[shortlabels]{enumitem}

\usepackage{color}
\usepackage[dvipsnames]{xcolor}
\definecolor{darkblue}{rgb}{0.1,0.1,.7}
\definecolor{purple}{rgb}{0.6,0,0.6}
\definecolor{orange}{rgb}{0.9,0.6,0}
\usepackage[colorlinks, linkcolor=darkblue, citecolor=darkblue, urlcolor=darkblue, linktocpage]{hyperref} 
\usepackage[square, comma, compress,numbers]{natbib}
\usepackage{physics}
\usepackage{tensor}
\usepackage{psfrag}
\usepackage{footmisc}
\usepackage{url}
\usepackage{mathtools}
\usepackage{tikz}
    \usepackage{amssymb,amsfonts,amsmath}
    \usepackage{tkz-euclide}
        \usetikzlibrary{arrows,calc,patterns}
\usepackage{pgfplots}
\usepackage[]{amsmath}
\usepackage[]{graphicx}
\usepackage[]{latexsym}
\usepackage[utf8]{inputenc}
\usepackage{geometry}
\usepackage{amscd}
\usepackage[all,cmtip]{xy}
\usepackage{mathrsfs}
\usepackage[most]{tcolorbox}
\usepackage{tikz}
\usepackage{subcaption}

\numberwithin{equation}{section}
\newcommand{\be}{\begin{equation}}
\newcommand{\ee}{\end{equation}}

\usepackage{ulem} 

\newcommand{\bea}{\begin{align}}

\newcommand{\eea}{\end{align}}

\numberwithin{equation}{section}

\newcommand{\ii}{\mathrm{i}}

\allowdisplaybreaks
\renewcommand{\i}{{\rm i}}

\newcommand{\xvec}{\bold{x}} 
\newcommand{\txvec}{\widetilde{\bold{x}}} 

\newcommand{\xvecNUT}{\bold{x}_{\text{NUT}}} 
\newcommand{\xvecANUT}{\bold{x}_{\text{aNUT}}} 
\newcommand{\GammaNUT}{\Gamma_{\text{NUT}}} 
\newcommand{\GammaANUT}{\Gamma_{\text{aNUT}}}

\newcommand{\nut}{\text{NUT}}

\newcommand{\GammaBH}{\Gamma_{\text{BH}}} 
\newcommand{\deltaBH}{\delta_{\text{BH}}} 
\newcommand{\GammaBL}{\Gamma_{\text{BL}}} 
\newcommand{\GammaBR}{\Gamma_{\text{BR}}} 
\newcommand{\deltaBR}{\delta_{\text{BR}}} 
\newcommand{\deltaBL}{\delta_{\text{BL}}}

\renewcommand{\i}{{\rm i}}

\newcommand{\bast}{\boldsymbol{\ast}}

\newcommand{\tgamma}{\widetilde{\gamma}}

\newcommand{\barZ}{\overline{Z}} 
\newcommand{\barOmega}{\overline{\Omega}} 
\newcommand{\onebar}{{\bar{1}}}
\newcommand{\twobar}{{\bar{2}}}
\newcommand{\threebar}{{\bar{3}}}

\renewcommand{\=}{\;= \;}
\newcommand{\IIA}{{\text{IIA}}}

\def\<{\langle}
\def\>{\rangle}

\def\Q{\mathcal{Q}}

\begin{document}

\thispagestyle{empty}
\vspace*{2cm}
\begin{center}

{\LARGE\bfseries
Novel black saddles for $5d$ gravitational indices \\ \vskip0.2cm
and the index enigma}
\begin{center}
\vspace{1cm}
 {\bf Jan Boruch${}^1$, Roberto Emparan${}^{2,3}$, Luca V. Iliesiu${}^1$,
Sameer Murthy${}^4$
}
\\
\bigskip \rm
\bigskip
${}^1$Center for Theoretical Physics and Department of Physics, University of California, Berkeley, California 94720, USA
\\
${}^2$ Institució Catalana de Recerca i Estudis Avançats (ICREA),
 Passeig Lluis Companys, 23, 08010 Barcelona, Spain
 \\
${}^3$Departament de Fisica Quàntica i Astrofisica and
  Institut de Ciències del Cosmos,
 Universitat de Barcelona, Marti i Franquès 1, 08028 Barcelona, Spain
\\
${}^4$Department of Mathematics, King's College London, The Strand, London WC2R 2LS, UK
\rm
\end{center}
\vspace{1 cm}
{\bf Abstract }
\end{center}
\begin{quotation}
\noindent
{We construct a series of novel Euclidean multi-black-holes, black ring, black Saturn, and black lens solutions to $5d$ supergravity 
that contribute as saddle-points to the $5d$ gravitational supersymmetric index, either in asymptotically flat space or in asymptotically AdS$_3\times S^2$. All these solutions are supersymmetric, have finite temperature, and an appropriate angular velocity turned on that makes fermionic fields periodic around the thermal circle. They contribute either to the helicity supertrace of supergravity in $5d$ flat space 
or to the elliptic genus of a supergravity theory in AdS$_3 \times S^2$. Their on-shell actions are independent of temperature, as consistent with the computation of a protected index, and equal to the entropy of the corresponding extremal black object. 
Our construction relies on uplifting saddles that can be singular in  $4d$, but which are desingularized in~$5d$. The resulting saddles exhibit a novel ``index enigma'', not encountered in previous Lorentzian solutions. One example of this enigma is that, in the computation of the index in asymptotically flat space, less symmetric black ring saddles dominate over the contributions from $5d$ black holes. }

\end{quotation}

\setcounter{page}{0}
\setcounter{tocdepth}{2}
\setcounter{footnote}{0}
\newpage

\tableofcontents

\newpage

\section{Introduction}

Supersymmetric indices have long served as computable observables in string theory that help us understand the microscopic structure of black holes~\cite{Sen:1995in, Strominger:1996sh}. 
By their nature, these indices are protected from continuous deformations -- of the string coupling, or of the temperature at which the index is evaluated 
-- and therefore capture exact information about BPS spectra that may otherwise be inaccessible. 
While the computation of the index on the string theory side is usually fairly straightforward (at least in principle), 
it was not known for a long time how to compute these indices directly from gravity. 
Indeed, the index from string theory has historically been compared to the entropy of extremal black holes, which is only defined at zero temperature -- and has no a priori reason to correctly capture the difference between bosonic and fermionic black hole degeneracies that the index computes. 
More recently, this problem was overcome by realizing that finite-temperature rotating black hole solutions, with an appropriately tuned angular velocity,  can preserve supersymmetry and, in fact, serve as the saddle points of the supergravity path integral in computations of the index~\cite{Cabo-Bizet:2018ehj,Iliesiu:2021are}. We call such solutions of supergravity \textit{index saddles}.

Much of the early progress in this area focused on single-center black hole saddles \cite{Sen:2008vm,Dabholkar:2010rm,Dabholkar:2010uh,Iliesiu:2022kny,LopesCardoso:2022hvc}, 
which provide leading contributions to indices, both in asymptotically flat space and in asymptotically $\mathrm{AdS}$ spacetimes \cite{Cabo-Bizet:2018ehj,Iliesiu:2021are,Cassani:2019mms,Bobev:2019zmz,Bobev:2020pjk,Larsen:2021wnu,Hristov:2021qsw,Hristov:2022pmo,BenettiGenolini:2023rkq, H:2023qko,Anupam:2023yns,Boruch:2023gfn,Hegde:2023jmp,Chowdhury:2024ngg,Chen:2024gmc,Cassani:2024kjn,Hegde:2024bmb,Adhikari:2024zif,Heydeman:2024fgk,Boruch:2025qdq,Cassani:2025sim,BenettiGenolini:2025jwe,Bandyopadhyay:2025jbc}. 
However, oftentimes these are not the only relevant geometries that are needed to fully compute the index. 
Multi-center configurations \cite{Denef:2000nb,Bates:2003vx,Denef:2007vg}, as well as solutions with non-spherical horizon topology, 
such as black rings~\cite{elvang2004supersymmetric} 
and black lenses~\cite{Kunduri:2014kja}, 
can play an essential role in ensuring the consistency of index computations. 
In this paper, we develop a systematic construction of novel index saddles of five-dimensional supergravity for a variety of such black objects.

Building on the framework of uplifting four-dimensional index geometries to five dimensions~\cite{Boruch:2025qdq}, we demonstrate how families of either smooth or singular Euclidean $4d$ configurations give rise to smooth, supersymmetric solutions in $5d$. 
This procedure yields new classes of saddles, including multi-black-hole configurations, black rings, black Saturns, and black lenses, 
both in asymptotically flat space and in asymptotically $\mathrm{AdS}_{3}\times S^{2}$. 
We show that these saddles exist at finite temperature, and, just like the single-centered index solutions found in the last few years, they have angular velocities tuned to preserve supersymmetry and a temperature-independent on-shell action that is consistent with index computations. 
Our analysis clarifies the conditions under which apparently pathological $4d$ saddles de-singularize upon uplift, thus enlarging the landscape of relevant contributions to the index.\footnote{See \cite{Bandyopadhyay:2025jbc,Cassani:2025iix} for related recent discussions which highlighted the importance of such contributions in the context of index saddles.}

By computing the on-shell action of all the saddles, we discover a novel index enigma, in which less symmetric saddles dominate in the path integral computation of the index over more symmetric saddles. 
While at first sight this seems analogous to the ``entropy enigma'' that was previously discussed in the literature~\cite{Denef:2007vg}, 
in numerous cases, the index enigma makes for a sharp comparison which does not have a good Lorentzian counterpart. 
This is because the computation of the index forces us to work in the grand-canonical ensemble, 
in which at least one of the angular velocities of the black objects is fixed to an appropriate value by the boundary conditions of the fields. 
This ensemble implements a sum over all saddles of varying angular momenta with that fixed angular velocity, 
and the leading contribution to the index is then given by a comparison between such saddles. 
In contrast, in Lorentzian signature, one typically compares the entropy between microcanonical phases that share the same charges and angular momenta (recall that in the supersymmetric Lorentzian solutions, 
the horizon angular velocity is zero), and there is often no clear meaning to a comparison of solutions with different charges and spins.

As a new example of this index enigma, we find that in $5d$ asymptotically flat space, the black ring saddles are more dominant in the path integral than the black hole saddles.\footnote{In Lorentzian signature, the supersymmetric black ring and black hole cannot, in general, be compared, since they carry different angular momenta: self-dual for the black hole and strictly non-self-dual for the black ring~\cite{elvang2004supersymmetric}.
} While this appears enigmatic in asymptotically flat space, by comparing saddles with the same choice of charges in asymptotically AdS$_3 \times S^2$, we arrive at a much more pedestrian conclusion: the BTZ black hole, which has the same charges as the black ring in flat space, dominates over the black hole with $S^3$ horizon in AdS$_3 \times S^2$. Put poetically, one spacetime's enigma is another spacetime's obvious solution.

The exploration of these more exotic saddles is motivated by microscopic calculations that suggest that the index encodes not only single-centered black hole states but also multi-centered bound states and horizonless configurations. 
For example, in $\mathcal{N}=4$ supergravity in four and five dimensions, the helicity supertrace has a generating function that is related to the Igusa cusp form \cite{Dijkgraaf:1996it,Gaiotto:2005gf,Shih:2005qf,Pioline:2005vi,David:2006yn,Castro:2008ys}, a Siegel modular form that was shown to include multi-center contributions from small black 
holes~\cite{Sen:2007vb,Dabholkar:2007vk,Cheng:2007ch,Dabholkar:2012nd}.\footnote{More general multi-center configurations do not contribute to the index in $\mathcal{N}=4$ supergravity and hence are not seen in the corresponding microscopic formula~\cite{Dabholkar:2009dq}.} 
In more general theories of~$\mathcal{N}=2$ supergravity, as is known from an analysis of gravitational solutions in Lorentzian signature, 
and as we now confirm from the saddles of the index, black rings, black Saturns, or multi-center black holes sometimes provide the leading contribution to the path integral in the $G_N^{-1}$ expansion.
Therefore, a complete understanding of the index requires incorporating all geometries consistent with supersymmetry and finite action. 
The results presented here extend the catalog of known index saddles and point toward a more comprehensive understanding of supersymmetric indices in higher-dimensional gravity.

 By systematically classifying and analyzing these index solutions, we aim to bridge the gap between microscopic index calculations 
and the semiclassical gravitational saddles that realize them. 
To be clear, the fact that different black objects, such as bound states of black holes, contribute to the microscopic index is not new: earlier work established it using a clever combination of semiclassical pictures and Lorentzian Hilbert spaces methods~\cite{Manschot:2010qz,Manschot:2011xc}.
In that approach, the different types of black objects are taken to be different ``molecules" in the same ensemble, to first approximation, and one subsequently attributes entropy to the molecules and counts them separately. 
The new point of view that we are advocating here uniformizes the treatment of all saddles: the simple statement is that many different index saddles, 
with possibly different horizon topologies, contribute to the same path integral for a gravitational index that is a priori defined. Each saddle contributes to the gravitational index through its Gibbons-Hawking on-shell action. 
This approach also paves the way towards exact computations of the index from the supergravity path integral, by including quantum corrections at each saddle.

The structure of the paper is as follows. In section~\ref{sec:General_construction} we review the $4d$/$5d$ uplift formula used to obtain the five-dimensional index saddles. We discuss the generalized construction of index saddles for singular four-dimensional geometries and show how they lead to smooth five-dimensional index saddles. In section~\ref{sec:index-saddles_flatspace} we apply this to find a general family of index saddles in five-dimensional asymptotically flat space. We analyze in detail the moduli space of these saddles, focusing on black holes, black rings, black Saturns, and a bound state of a black hole and a black lens. In section~\ref{sec:indices-in-ads3-s2} we revisit this discussion for saddles with AdS$_3 \times S^2$ asymptotics. We present in detail the cases of black holes, black lenses, and their possible bound states. In section~\ref{sec:moduli_space_discussion} we discuss general aspects of the moduli spaces of smooth $5d$ index saddles obtained from our construction. Finally, in section~\ref{sec:discussion} we summarize and discuss future directions. In particular, we comment on a puzzling temperature dependence of the indices, arising from the disappearance of real saddles from the moduli space for sufficiently high temperatures.

\section{General construction of index saddles in $5d$ supergravity}
\label{sec:General_construction}

In this section, we present the method we employ to construct five-dimensional saddles for gravitational indices. 
A general family of such saddles can be obtained via the $4d$/$5d$ lift \cite{Elvang:2005sa,Bena:2005ni,Gaiotto:2005gf,Behrndt:2005he,Gaiotto:2005xt, Banerjee:2011ts}, which relates solutions of four-dimensional supergravity, obtained from compactification of type IIA supergravity on a Calabi-Yau manifold, 
to solutions of five-dimensional supergravity obtained from compactifying eleven-dimensional supergravity on the same Calabi-Yau manifold. 

In the context of gravitational indices, this route was previously followed in \cite{Boruch:2025qdq} to construct black hole and black ring index saddles in asymptotic $\mathbb{R}^4 \times S^1$ and AdS$_3 \times S^2$, respectively. 
Here, our aim is to construct more general five-dimensional index saddles, including configurations with multiple black holes, black lenses, and black rings. 
As we will see, in some cases this requires us to consider a specific family of singular four-dimensional configurations that become desingularized upon uplift to five dimensions. 

We begin in section \ref{sec:uplfiting_4d_to_5d} by reviewing the $4d$/$5d$ uplift formula and laying out the explicit conventions that we follow throughout the paper. 
In section \ref{sec:4d-to-5d-smoothness}, we explain, from the five-dimensional perspective, why the general family of four-dimensional index saddles~\cite{Boruch:2025biv} must include singular configurations to account for cases such as black rings in $5d$ asymptotically flat space. 
Finally, in section \ref{sec:three_centers_general}, we revisit the $4d$ analysis of smoothness for a representative singular three-center configuration and clarify the conditions under which it yields a smooth five-dimensional solution.

\subsection{Uplifting $4d$ saddles to $5d$}
\label{sec:uplfiting_4d_to_5d}

We are interested in constructing index saddles of five-dimensional supergravity coupled to $n_V-1$ ($n_V=1,2,\dots$)
vector multiplets~\cite{Bodner:1990zm,Cadavid:1995bk,Kraus:2005gh,Larsen:2006xm,Looyestijn:2010pb}. 
This theory can be viewed as a low-energy effective description of M-theory compactified on a general Calabi-Yau manifold.
A large family of solutions, containing a $U(1)$ Killing vector along the M-theory circle, can be directly related to solutions of $\mathcal{N}=2$ four-dimensional supergravity coupled to $n_V$ vector multiplets through a $4d$/$5d$ lift~\cite{Behrndt:2005he,Gaiotto:2005gf,Gaiotto:2005xt,Elvang:2005sa,Bena:2005ni, Banerjee:2011ts} as follows,
\begin{align}
\frac{\dd s_{5d}^2}{\ell_5^2} &\= (2\widetilde{V}_\IIA)^{2/3} \, (\dd\psi+A^0)^2 + (2\widetilde{V}_\IIA)^{-1/3} 
\, {\dd s^2_{4d}} \,, \label{eq:5d_uplift_formula_1}
\\  
{A^A_{5d}}&\= -A^A + \Re t^A \, (\dd\psi + A^0) \,, 
\qquad 
Y^A\= \frac{\Im t^A}{\widetilde{V}_\IIA^{1/3}}
\,. \label{eq:5d_uplift_formula_2}
\end{align}
Here $\psi \sim \psi + 4\pi$ and, in terms of fields of four-dimensional theory, $\widetilde{V}_{\text{IIA}}$ denotes the dimensionless Calabi-Yau volume measured in string units,
$A^{0}$ is the graviphoton of the gravity multiplet and $(t^A \equiv B^A + \i J^A ,A^A)_{A=1,\dots,n_V}$ denote complex scalars and gauge fields of the $n_V$ vector multiplets of the four-dimensional theory. 
To construct the five-dimensional saddles we will apply the $4d$/$5d$ lift directly to supersymmetric index saddles of $\mathcal{N}=2$ four-dimensional supergravity\footnote{For a more detailed exposition of four-dimensional $\mathcal{N}=2$ supergravity in this context, see e.g.~\cite{Mohaupt:2000mj,Boruch:2023gfn}.} in asymptotically flat space \cite{Boruch:2023gfn,Boruch:2025biv} of the following sort, 
\begin{align}
\dd s^2_{4d} &\= \frac{1}{\Sigma(H)} (\dd t + \omega_E)^2 + \Sigma(H) \dd x^m\dd x^m \, , 
\qquad 
\bast \, \dd \omega_E \= \ii \langle \dd H , H \rangle \,,
\label{eq:4d_attractor_saddle_metric} 
\\
A^{\alpha} & \= -\ii I^{\alpha \beta} \partial_{H^\beta} \log (\Sigma) 
\left( \dd t+ \omega_E \right)
+ \mathcal{A}_d^\alpha \,, 
\qquad 
\dd \mathcal{A}_d^\alpha \= \bast \, \dd H^\alpha 
\,,
\label{eq:explicit_gauge_field_A0}
\end{align}
where we work with saddles compactified in the Euclidean time $t \sim t+\beta$.
Here, 
\begin{equation}
    H\= (H^0 , H^A , H_A , H_0)
\end{equation}
is a vector of harmonic functions on $\mathbb{R}^3$ base space 
which transform as a symplectic vector under electric-magnetic duality. For two symplectic vectors $A,B$, we define the duality invariant product as
\be 
\langle A , B \rangle \= A^I B_I - A_J  B^J \=  
A^\alpha I_{\alpha \beta} B^\beta \, , 
\qquad
I_{\alpha \beta} \= 
\begin{pmatrix}
0 &0 & 0 & 1 \\
0 & 0 & 1 & 0 \\
0 & -1 & 0 & 0 \\
-1 & 0 & 0 & 0 
\end{pmatrix}
\,.
\ee 
Scalar fields $t^A$ can also be conveniently repackaged into a symplectic vector, known as the normalized period vector,
\be 
\Omega \= 
\frac{1}{\sqrt{\frac{4}{3} D_{ABC} J^A J^B J^C}}
 \left(-1 , -t^A , - \frac{t_A^2}{2} , 
\frac{t^3}{6}\right),
\quad
t_A^2 \; \equiv \;  D_{ABC} t^A t^B , \quad 
t^3 \; \equiv  \; D_{ABC} t^A t^B t^C \, . 
\label{eq:normalized_period_vector}
\ee
In this paper, we will work with a two-derivative theory specified by a cubic prepotential \be 
F = \frac{1}{6} D_{ABC} \frac{X^A X^B X^C}{X^0}
\ee 
which, in turn, is given in terms of projective coordinates on the scalar moduli space $t^A = \frac{X^A}{X^0}$.\footnote{For details, see \cite{Boruch:2023gfn,Boruch:2025qdq}.}

The function $\Sigma(H)$ is explicitly determined for each Calabi-Yau manifold as follows~\cite{Shmakova:1996nz}, 
\begin{align}
\Sigma(H) &\= \sqrt{\frac{\Q_D^3-L_D^2}{(H^0)^2}} \,, 
\qquad \Q_D^{3/2} \= \frac{1}{3} D_{ABC} \, y^A y^B y^C \,,
\label{eq:Q_function_Shmakova}
\\ 
L_D&\= 
-
H_0 (H^0)^2 + \frac{1}{3} D_{ABC} H^A H^B H^C - H^A H_A H^0 \,. 
\label{eq:L_function_Shmakova}
\end{align}
Here $y(H)$ are functions of $H$ determined from the algebraic set of equations 
\be 
D_{ABC}\, y(H)^A y(H)^B\= D_{ABC} H^A H^B - 2 H_C H^0
\label{eq:y_function_Shmakova}
\,,
\ee
where $D_{ABC}$ denotes the triple-intersection number of a given Calabi-Yau manifold. The function $\Sigma$ is often referred to as the entropy function, as it determines the area of the extremal black hole through 
\be 
S_{\text{ext}} \= \pi \Sigma(\Gamma),
\label{eq:extremal_entropy_in_Sigma}
\ee
for a black hole with total charge 
\begin{equation}
    \Gamma = (P^0 , P^A, Q_A , Q_0), 
\end{equation}
where $P^0,P^A$ denote magnetic charges and $Q_A,Q_0$ denote electric charges. The existence of the Killing spinors is guaranteed by the generalized attractor equations \cite{Ferrara:1995ih,Ferrara:1996dd,Denef:2000nb}
\be 
H(\xvec)\= \ii (\overline{Z}_* (H) \Omega_* (H) - Z_* (H) \overline{\Omega}_* (H)) ,
\ee
where $\Omega_* (H) \equiv \Omega (t(H))$ encodes the solution to the above equations for the scalar fields in terms of the harmonic functions $H$, and $Z_*(H) \equiv \langle H, \Omega_* (H) \rangle$, $\overline{Z}_*(H) \equiv \langle H, \overline{\Omega}_* (H) \rangle$.

The four-dimensional black hole index saddles \cite{Boruch:2023gfn,Boruch:2025biv} are specified by choosing a set of monopole charges $\Gamma_i$, 
base-space positions of north/south poles of each black hole $(\xvec_i , \txvec_i)$, as well as asymptotic boundary conditions for the scalar fields $\Omega_\infty \equiv \Omega(t_\infty)$. With this data, the harmonic functions take the form 
\be 
H(x) \= h + \sum_{i=1}^N \frac{\gamma_i}{|\xvec - \xvec_i|} 
+ \sum_{i=1}^N \frac{\tgamma_i}{|\xvec - \txvec_i|} ,
\qquad 
h \= \i (e^{-\i \alpha_\infty} \Omega_\infty - e^{\i \alpha_\infty} \overline{\Omega}_\infty ) ,
\label{eq:4d_attractor_saddle_harmonic_function}
\ee
with $\i \alpha_\infty \equiv \arg(\langle \Gamma , \Omega_\infty \rangle)$, $\Gamma = \sum_{i=1}^N \Gamma_i$, and the coefficients of the poles determined through their corresponding monopole charges by the new attractor mechanism~\cite{Boruch:2023gfn},
\begin{align}
\gamma_i \; \equiv \; \frac{\Gamma_i}{2} + \i \delta_i  
 \= i \barZ_*(\Gamma_i) \Omega_*(\Gamma_i) \,, 
\qquad
\tgamma_i \; \equiv \; \frac{\Gamma_i}{2} - \i \delta_i  
\= - i Z_*(\Gamma_i) \barOmega_*(\Gamma_i)
\,.
\label{eq:4d_attractor_saddle_charges}
\end{align}
We see that, while $\Gamma_i$ is the total (i.e., monopole) charge of the system, $\delta_i$ are (imaginary) dipole charges. The smoothness of the four-dimensional saddles is ensured by imposing the removability of Dirac-Misner strings for the one-form $\omega_E$. This constrains the allowed base space positions of the poles $(\xvec_i, \txvec_i)$ through a set of equations 
\begin{align}
i \= 1, \dots , N, \qquad 
\i \langle \gamma_i , H(\xvec_i)  \rangle \= \frac{\beta}{4\pi} 
\,, \qquad 
\i \langle \tgamma_i , H(\txvec_i)  \rangle \= - \frac{\beta}{4\pi}  \,.
\label{eq:multicentered_regularity_condition}
\end{align}

From these four-dimensional index saddles, the uplifted M-theory saddles with $\mathbb{R}^3 \times S^1 \times S^1 \times \text{CY}_3$ asymptotics then directly follow from the $4d$/$5d$ uplift \eqref{eq:5d_uplift_formula_1}, \eqref{eq:5d_uplift_formula_2}. For many purposes, it is useful to simplify the form of the metric by using the explicit expressions \eqref{eq:Q_function_Shmakova}, \eqref{eq:L_function_Shmakova}, and the volume of the Calabi-Yau manifold  
\be 
\widetilde{V}_{\text{IIA}} 
\;\equiv \; \frac{1}{6} D_{ABC} J^A J^B J^C
\= \frac{1}{2} \left(
\frac{\Sigma}{\Q_D}
\right)^3  ,
\label{eq:CY_volume_space_dependent}
\ee
This allows us to rewrite the full five-dimensional metric as
\begin{align}
\label{eq:5dmetricQL_Euclidean}
\dd s^2_{5d} &\= 
\frac{(H^0)^2}{\Q_D^2} \left(
\dd t + \omega_E 
+ \ii \frac{L_D}{(H^0)^2}(\dd \psi + \mathcal{A}_d^0) \right)^2 
+\frac{\Q_D}{H^0} \dd s^2_{\text{TN}} \,,  
&
\bast \, \dd \omega_E &\= \ii \langle \dd H ,H \rangle \,,
\end{align}
with
\begin{align}
\dd s^2_{\text{TN}}
&\=
\frac{1}{H^0} 
(\dd \psi + \mathcal{A}_d^0)^2 
+ H^0 \dd x^m \dd x^m 
\,, 
& \bast \, \dd \mathcal{A}_d^0 
&\= \dd H^0 \,,\label{eq:TN}
\end{align}
which directly shows the special role played by the D6 charges $P^0_i$. For a single four-dimensional black hole saddle, these uplifted solutions were studied in \cite{Boruch:2025qdq}. Here, our goal is to extend such uplifts to multicentered black holes, as well as to objects with zero extremal horizon area, such as one-charge and two-charge black holes. In particular, the one-charge case requires us to extend the most general form of the finite temperature harmonic functions \eqref{eq:4d_attractor_saddle_harmonic_function}, as we explain in the following section.

\subsection{From $4d$ singular saddles to $5d$ smooth saddles}
\label{sec:4d-to-5d-smoothness}

Some of the solutions that we are interested in exhibit $4d$ singularities at the poles, rather than regular horizons. The most relevant case is that of poles with only D6 charge, i.e., $\Gamma=( \Gamma^0=1,0,0,0)$, which, as we mentioned, is singled out in the $4d$/$5d$ uplift. These poles appear in black rings and black lenses.

In the formalism introduced above, the solution for an extremal D6-brane has $\omega_E=0$ and $\Q_D=H^0$, with
\begin{equation}\label{eq:KKm_4d_H0}
    H^0\= h^0 +\frac{\Gamma^0}{|\xvec|}\,.
\end{equation}
Although the four-dimensional metric
\begin{equation}\label{eq:KKm_4d_metric}
    \dd s^2_{4d} \=\frac{1}{\sqrt{H^0}}\dd t^2
    +\sqrt{H^0}\, \dd x^m \dd x^m
\end{equation}
has a curvature singularity at $|\xvec|=0$, it has been long known \cite{Sorkin:1983ns,Gross:1983hb} that its $5d$ lift
\begin{equation}\label{eq:KKmonopole}
    \dd s^2_{5d} \= \dd t^2+\dd s^2_{\text{TN}}\,,
\end{equation}
with a factor of the four-dimensional Taub-NUT metric \eqref{eq:TN}, has a much milder $A_{\Gamma^0-1}$ conical singularity, and is actually smooth for unit D6-charge, $\Gamma^0=1$.

There also exist `purely fluxed' solutions describing a D6-brane with D4, D2, and D0 charges. In these configurations, a gauge flux is turned on on the D6 worldvolume, inducing D2 charge through Chern–Simons couplings and a D0 charge that, in the M-theory uplift, corresponds to the Poynting momentum generated by the M2 and M5 charges. The D2 and D0 charges are therefore not independent parameters in this D6-D4-D2-D0 system, which has neither a horizon nor an associated entropy. The solution can be obtained via a large gauge transformation of \eqref{eq:KKmonopole} in M-theory,\footnote{The large gauge transformation in M-theory, which leaves both the metric and the entropy invariant, acts on the harmonic functions as $H^0 \to H^0$, $H^A \to  H^A - H^0 k^A$, $H_A \to  H_A - D_{ABC} H^B k^C + \frac{H^0}{2} D_{ABC} k^B k^C$, $H_0  \to H_0 + k^A H_A - \frac{1}{2}D_{ABC} H^A k^B k^C + \frac{H^0}{6} D_{ABC} k^A k^B k^C$ \cite{Cheng:2006yq,deBoer:2008fk}.\label{foot:gaugeM}} where its geometry is smooth. The main features of our analysis below extend to this case as well.

Returning to the ansatz \eqref{eq:4d_attractor_saddle_harmonic_function}, we see the phenomenon of center splitting: a single center of charge $\Gamma_i$ is replaced,  in finite-temperature index geometries, by two centers of charge $\Gamma_i/2\pm i\delta_i$.
This splitting correctly produces the index saddles for extremal black holes, but it is not obvious whether it should also be applied to the poles with $\Gamma=( \Gamma^0=1,0,0,0)$. As we explained above, these poles are present in black rings and black lenses, and in four dimensions, they appear as singularities in the geometry.

A similar question arose in the study of indices for two-charge configurations, i.e., “small black holes.’’ For these cases, refs.~\cite{Chowdhury:2024ngg,Chen:2024gmc} argued that the singular poles should be split to obtain index saddles. They constructed split-center index geometries with smooth Einstein-frame metrics and Euclidean horizons of finite area, though one scalar diverges at the poles. In the heterotic frame, where this scalar is the string coupling, ref.~\cite{Chen:2024gmc} argued that stringy higher-curvature effects resolve the issue. The question we ask now is whether one-charge poles, as appear in D6-branes, admit analogous split-center index geometries.

To address this question, rather than use the general multi-center solutions, we attempt to construct an index solution for a D6-brane starting with an appropriate Euclidean rotating black hole and imposing the condition $\beta\Omega = 2\pi i$ to realize $(-1)^F$ in the gravitational path integral. 

The relevant solution here is a non-extremal, rotating black hole with D6 charge, which has the Euclideanized metric
\begin{align}\label{eq:rotbh}
    \dd s^2\=\frac{1}{\sqrt{H^0}}\left(\dd t +\mathcal{A}\right)^2
    +\sqrt{H^0}\left( \Sigma\left(\frac{\dd r^2}{\Delta}+\dd \theta^2\right) +\Delta \sin^2\theta\,\dd \phi^2\right)
\end{align}
with
\begin{align}
    \Delta &\=r^2-2mr-\textsf{a}^2\,,\\
    \Sigma &\=r^2-2mr-\textsf{a}^2\cos^2\theta\,,\\
    \mathcal{A}&\= i \frac{2m \textsf{a} \cosh \delta }{G} r \sin^2\theta \,\dd \phi\,,\label{eq:calA}\\
    H^0&\=\frac{r^3 (r+2m \sinh^2\delta)-2\textsf{a}^2\cos^2\theta\left( r^2 + m r \sinh^2\delta\right) + \textsf{a}^4 \cos^4\theta}{\Sigma^2}\;.\label{eq:H0}
\end{align}
The parameters $m$, $\delta$, and $\textsf{a}$ control non-extremality, charge boost, and imaginary rotation. This metric, and the ensuing analysis, agree with setting one of the charges to zero in the two-charge solutions of \cite{Chowdhury:2024ngg,Chen:2024gmc}, but the absence of the second charge alters the conclusions.\footnote{In particular, \eqref{eq:rotbh} follows from Ref.~\cite{Chen:2024gmc}, Secs.~2 and 3, by taking $\upalpha = \upbeta \equiv \delta$, which yields $Q_L = Q_R$, i.e., only momentum charge $n$, and no winding, $w=0$. In the solutions below, $q=n$. Choosing $\upalpha = -\upbeta$ leaves instead winding as the only charge.}

To get oriented, set $\textsf{a}=0$ and take $m \to 0$, $\delta \to \infty$ while holding
\begin{equation}\label{eq:mdeltalimit}
\frac{m}{2} e^{2\delta} \; \equiv \; q
\end{equation}
finite. Then
\begin{equation}
   H^0 \= 1 + \frac{q}{r}\,, 
\end{equation}
and thus we recover the single-center solution \eqref{eq:KKm_4d_H0} for a D6-brane, with $h^0=1$ and $\Gamma^0=q$.

Now we attempt to build a finite-temperature index geometry using \eqref{eq:rotbh}. Its physical magnitudes are
\begin{eqnarray}
    &G_4 M \= \frac{m}{4}(3+\cosh 2\delta)\,,
    &G_4 Q \= \frac{m}{4}\sinh 2\delta\,,\\
    &G_4 J \= i m \textsf{a} \cosh \delta\,,
    &S \= \frac{2\pi m \cosh\delta}{G_4}\left( m + \sqrt{m^2+\textsf{a}^2}\right)\,,\\
    &\beta\=4\pi m \cosh\delta \left(\frac{m}{\sqrt{m^2+\textsf{a}^2}}+1\right)\,,
    &\beta \Omega \= 2\pi i\frac{\textsf{a}}{\sqrt{m^2+\textsf{a}^2}}\,.
\end{eqnarray}
To impose $\beta\Omega = 2\pi i$ we must take $m \to 0$ (or, possibly, $\textsf{a} \to \infty$, but this does not yield a well-defined limit).
Again, to keep a non-zero and finite charge, we also send $\delta \to \infty$ with \eqref{eq:mdeltalimit}
finite. This limit satisfies the BPS condition $M = Q$, so for $\textsf{a} \neq 0$ we may expect to obtain the desired index geometry.

However, even if $\textsf{a} \neq 0$ we find that $\beta \to 0$  in this limit,\footnote{This feature already appeared in \cite{Chen:2023mbc}. The uplift of \eqref{eq:rotbh} to five dimensions yields a neutral rotating black hole, and one could directly extend the analysis of \cite{Chen:2023mbc} to compute $\mathrm{Tr}\left[(-1)^F e^{-\beta H}\right]$ for the non-BPS system.}
and also both the entropy and the angular momentum vanish. The latter is troublesome, since our aim was to use the spin-statistics relation to compute the index. The vanishing entropy also suggests that the geometry is singular, and not a smooth index saddle.

The nature of this solution is clarified by rewriting it in a different form. In the supersymmetric limit, \eqref{eq:calA} and \eqref{eq:H0} reduce to
\begin{align}
\mathcal{A} \= 0\,, \qquad
H^0 \= 1 + \frac{q/2}{r+\textsf{a}\cos\theta}
+ \frac{q/2}{r-\textsf{a}\cos\theta}\,.
\end{align}
The fact that  $\mathcal{A}=0$ shows that the geometry is static, as already suggested by the vanishing of $J$. Furthermore, the form of $H^0$
reveals that this is a two-center solution (as expected), with poles at $r = \pm \textsf{a}\cos\theta$.
Indeed, under the coordinate change $(r,\theta,\phi) \to (x,y,z)$,
\begin{align}
x + i y &\= \sqrt{r^2 - \textsf{a}^2}\,\sin\theta\, e^{i\phi}\,, \qquad
z \= r\cos\theta\,,
\end{align}
the metric \eqref{eq:rotbh} becomes of the form \eqref{eq:KKm_4d_metric}
with
\begin{align}
H^0 \= 1 + \frac{\Gamma^0/2}{|\mathbf{x} - \mathbf{x}_N|}
+ \frac{\Gamma^0/2}{|\mathbf{x} - \mathbf{x}_S|}\,, \qquad
\mathbf{x}_{N,S} \= (0,0,\pm \textsf{a})\,.
\end{align}
where $\Gamma^0=q$. Thus, the solution is only a trivial split
into two centers of charge $\Gamma^0/2$ and no dipole $\delta^0$ nor rotation. These centers have the same zero-area, infinite-temperature singularity (with divergent curvature and scalars) as the initial single-center solution.

Such a split does not yield good index saddles. In particular, it must not be applied to a D6-brane with $\Gamma^0=1$: the half-split centers uplift to singular five-dimensional geometries. The correct index geometry is the single-center one, \eqref{eq:KKmonopole}, which can clearly have arbitrary Euclidean time periodicity and thus is suitable for computing $\mathrm{Tr}\left[(-1)^F e^{-\beta H}\right]$.

This contrasts sharply with the two-charge case \cite{Chowdhury:2024ngg,Chen:2024gmc}. There, if one charge (say, winding $w$) is small but nonzero, higher-derivative corrections (eq.~5.19 of \cite{Chen:2024gmc}) give the expected behavior: $\beta$ grows large rather than vanishing. But the $w \to 0$ limit is discontinuous, and one-charge configurations cannot be cured by such corrections.

We conclude that, when constructing general five-dimensional index saddles, poles corresponding to one-charge objects of vanishing extremal area---such as D6 centers (in $4d$) or NUT centers (in $5d$) with $\Gamma_{\text{NUT}}=( \Gamma^0=1,0,0,0)$---should not split into north-pole/south-pole pairs at finite temperature~\cite{Bandyopadhyay:2025jbc,Cassani:2025iix}. This contrasts with the centers in \eqref{eq:4d_attractor_saddle_harmonic_function}, which represent finite-area black holes~\cite{Boruch:2023gfn,Boruch:2025biv}.

\subsection{General $5d$ attractor saddles}
\label{sec:three_centers_general}

The analysis of the previous subsection motivates us to consider a more general family of harmonic functions
\be 
H(x) \= h + \sum_{i=1}^{N_{NUT}}\frac{\Gamma_{\text{NUT},i}}{|\xvec-\xvec_{\text{NUT},i}|}
+
\sum_{i=1}^{N_{BH}}
\left(\frac{\gamma_i}{|\xvec - \xvec_{i}|}
+ \frac{\tgamma_i}{|\xvec - \txvec_i|} 
\right)
,
\label{eq:5d_attractor_saddles_harmonic}
\ee
for total charge splittings such that $\Sigma(\Gamma_{\text{NUT},i})=0$ and $\Sigma(\gamma_i + \tgamma_i) >0$.
Here, the north and south poles are paired up and connected by a removable Dirac-Misner string to form a black hole, whilst the remaining poles $\xvec_{\text{NUT},i}$ stay unpaired.

We now have two choices of what we can do with Dirac-Misner strings coming out of the unpaired poles. 
We could allow them to have removable Dirac-Misner strings that go off to infinity---this leads to a solution in which the $4d$ metric has Taub-NUT asymptotics (where the~$S^1$ is non-trivially fibered on $\mathbb{R}^3$), and hence it is asymptotically locally flat. 
Alternatively, we could impose an integrability condition that would guarantee that there is no Dirac-Misner string 
coming out of the remaining pole~\cite{Warner:2019jll}, and correspondingly the $4d$ metric is asymptotically globally~$\mathbb{R}^3 \times S^1$, 
and hence asymptotically flat. 
We will concentrate on the latter case, as we are interested in asymptotically flat $4d$ metrics. The resulting smoothness conditions then take the form
\begin{align}
\label{eq:5dsmoothness}
\i \langle \gamma_i , H(\xvec_i) \rangle &\= \frac{\beta}{4\pi} , 
\qquad 
\i \langle \tilde \gamma_i , H(\tilde \xvec_i) \rangle \= -\frac{\beta}{4\pi} , \qquad i \= 1, \dots, N_{BH} \,,
\\
\langle \Gamma_{\text{NUT}, i} , H(\xvec_{\text{NUT}, i}) \rangle &\= 0, \qquad i \= 1, \dots, N_{NUT} \,. 
\label{eq:5dintegrability}
\end{align}

Clearly, saddles with zero-area centers are singular in 
$4d$. For suitable choices of charges, however, these singularities are resolved upon lifting to $5d$: the apparent $4d$ pathology reflects the contractibility of the M-theory circle, whose size depends on four-dimensional scalars.
In such cases, the singular $4d$ index saddle corresponds to a smooth $5d$ one. To guarantee this, we require that no singularities beyond the scalar divergence at the zero-area poles occur in $4d$. This is precisely enforced by \eqref{eq:5dsmoothness} and \eqref{eq:5dintegrability}.

We now analyze an example of such a configuration and show how it can indeed lead to smooth $5d$ saddles.
The simplest non-trivial case we can consider is that of three-center saddles (for the cases with a single $4d$ black hole, we denote $\xvec_1 \equiv \xvec_N, \gamma_1 \equiv \gamma_N $, $\txvec_1 \equiv \xvec_S, \tgamma_1 \equiv \gamma_S$)
\be 
H(x) \= h + \frac{\GammaNUT}{|\xvec-\xvecNUT|} + \frac{\gamma_N}{|\xvec - \xvec_{N}|}
+ \frac{\gamma_S}{|\xvec - \xvec_{S}|} \;,
\label{eq:three_center_harmonic}
\ee
consisting of a single NUT pole with $\Sigma(\GammaNUT)=0$ and a north-south pole pair of a black hole saddle with charge $\Gamma= \gamma_N + \gamma_S$ such that $\Sigma(\Gamma)>0$. 

The smoothness conditions \eqref{eq:5dsmoothness}, \eqref{eq:5dintegrability}, impose
\begin{align}
\i \langle \gamma_N , H(\xvec_N) \rangle \= \frac{\beta}{4\pi} \,,
\qquad 
\i \langle \gamma_S , H(\xvec_S) \rangle \= -\frac{\beta}{4\pi} \,,
\qquad
\langle \GammaNUT , H(\xvecNUT) \rangle \= 0 \,. 
\end{align}
Let us now try to solve the above equations. Explicitly we have
\begin{align} \label{eq:3centers}
\i \langle \gamma_N , h \rangle + 
\frac{\i \langle \gamma_N , \gamma_S \rangle}{|\xvec_N - \xvec_S|} 
+ 
\frac{\i \langle \gamma_N , \GammaNUT \rangle}{|\xvec_N - \xvecNUT|} 
&\= 
\frac{\beta}{4\pi} 
\,, \\
\i \langle \gamma_S , h \rangle + 
\frac{\i \langle \gamma_S , \gamma_N \rangle}{|\xvec_N - \xvec_S|} 
+ 
\frac{\i \langle \gamma_S , \GammaNUT \rangle}{|\xvec_S - \xvecNUT|} 
&\= 
- \frac{\beta}{4\pi}\, , 
\\
\langle \GammaNUT , h \rangle 
+ \frac{\langle \GammaNUT , \gamma_N \rangle}{|\xvec_N - \xvecNUT|}
+  \frac{\langle \GammaNUT , \gamma_S \rangle}{|\xvec_S - \xvecNUT|}
& \= 0 
\,.
\end{align}

For three centers, we expect to have $9-6=3$ independent parameters that describe the configuration after fixing rotations and translations of the solutions. 
The conditions~\eqref{eq:3centers} are 6 real equations coming from 3 complex equations, and the above expectation will be realized if only three of the equations are linearly independent. 
This can be explicitly checked as follows. 
Firstly, because $\langle \Gamma_{\text{total}} , h \rangle = 0$,  the sum of the three equations above gives a tautological $0=0$. 
So we can safely ignore one of the complex equations, say the second one. 
We are therefore left with the following four real equations so far,
\begin{align}
\langle \Gamma , h \rangle 
\;+\; 
\frac{\langle \Gamma , \GammaNUT \rangle}{|\xvec_N - \xvecNUT|} & \= 0 \, , 
\label{eq:three-centers_regularity_eq1}
\\ 
\langle h,\delta \rangle \;+\; \frac{\langle \Gamma , \delta \rangle}{|\xvec_N - \xvec_S|} \;+\; 
\frac{\langle \GammaNUT , \delta \rangle}{|\xvec_N - \xvecNUT|} &\= \frac{\beta}{4\pi} \,,
\label{eq:three-centers_regularity_eq2}
\\
\langle \GammaNUT , h \rangle 
\;+\; \frac{1}{2} \frac{\langle \GammaNUT , \Gamma \rangle}{|\xvec_N - \xvecNUT|}
\;+\;  \frac{1}{2} \frac{\langle \GammaNUT , \Gamma \rangle}{|\xvec_S - \xvecNUT|}
&\= 0 \,,
\label{eq:three-centers_regularity_eq3}
\\
\frac{\i \langle \GammaNUT , \delta \rangle}{|\xvec_N - \xvecNUT|}
\;-\; \frac{\i \langle \GammaNUT , \delta \rangle}{|\xvec_S - \xvecNUT|}
&\= 0 \,,
\label{eq:three-centers_regularity_eq4}
\end{align}
where we recall that~$\Gamma=\gamma_N+\gamma_S$ and $\delta=\dfrac{1}{2 i}(\gamma_N-\gamma_S)$.
The last equation \eqref{eq:three-centers_regularity_eq4} imposes that 
\be \label{eq:isosceles}
|\xvec_N - \xvecNUT| \= |\xvec_S - \xvecNUT| . 
\ee
Plugging this back into the above equations, we immediately see that \eqref{eq:three-centers_regularity_eq1} and \eqref{eq:three-centers_regularity_eq3} become identical after using $\langle \Gamma , h \rangle = - \langle \GammaNUT , h \rangle$, which is the third linearly dependent equation we were looking for. 
The remaining equations \eqref{eq:three-centers_regularity_eq1} and \eqref{eq:three-centers_regularity_eq2} can now be solved for the distances. They yield 
\begin{align}
|\xvec_N - \xvecNUT| &\= |\xvec_S - \xvecNUT| \= 
- \frac{\langle \Gamma , \GammaNUT \rangle}{\langle \Gamma, h \rangle} \,,
\\ 
|\xvec_N - \xvec_S|
&\=  \frac{\langle \Gamma, \delta \rangle}{\frac{\beta}{4\pi}- \langle h ,\delta \rangle - \frac{\langle \GammaNUT , \delta \rangle}{\langle \GammaNUT , \Gamma \rangle}
\langle \Gamma, h \rangle} 
\,.
\end{align} 
Having found solutions to the smoothness equations, we now need to make sure that the resulting values for the three distances can be properly embedded in $\mathbb{R}^3$.
To ensure this, we impose triangle inequalities.
From \eqref{eq:isosceles} we see that the triangle is isosceles with the symmetry~$\xvec_N \leftrightarrow \xvec_S$, and we should impose 
\be 
|\xvec_N - \xvec_S| \; \geq \; 0 \, , \qquad
|\xvec_N - \xvecNUT|+|\xvec_S - \xvecNUT| \= 2|\xvec_N - \xvecNUT| \; \geq \; |\xvec_N - \xvec_S|\,
. 
\ee
The first condition simply imposes that we must work at low enough temperature, i.e.,
\be 
\beta \; \geq \; \beta_{\text{cr},1} \,,
\qquad 
\beta_{\text{cr},1} \; \equiv\;
4\pi \left( \langle h ,\delta \rangle + \frac{\langle \GammaNUT , \delta \rangle}{\langle \GammaNUT , \Gamma \rangle}
\langle \Gamma, h \rangle \right) \,,
\label{eq:three-center_temperature_condition}
\ee
as is usual in attractor saddles \cite{Boruch:2023gfn}. 
The second condition leads to a constraint on the temperature
\be 
\beta \; \geq \; \beta_{\text{cr},2}, \qquad \beta_{\text{cr},2} \; \equiv \; \beta_{\text{cr},1} - 2\pi
 \langle \Gamma,\delta \rangle
 \frac{\langle \Gamma, h \rangle }{\langle \Gamma , \GammaNUT \rangle} 
\,.
\label{eq:three-center_temperature_condition_2}
\ee
Now, since for the $4d$ bound state to exist at extremality, we need 
\be 
- \frac{\langle \Gamma , \GammaNUT \rangle}{\langle \Gamma, h \rangle} \;\geq \; 0 \,,
\label{eq:three_center_distance_condition}
\ee
and $\langle \Gamma,\delta \rangle$ is the extremal entropy and hence positive, we see that the second condition is stronger.
The bottom line is that it is enough to impose the condition in \eqref{eq:three-center_temperature_condition} for the existence of
the bound state and, 
in particular, the low temperature condition of $\beta \geq \beta_{\text{cr},1}$ is always satisfied whenever a bound state makes sense. 
As we approach $\beta \to \beta_{\text{cr},2}$ from above, the triangle formed by three poles degenerates and becomes colinear, since 
\be 
\beta \to \beta_{\text{cr},2} \qquad 
\Rightarrow \qquad 
|\xvec_N - \xvec_S| \to |\xvec_N - \xvecNUT| + |\xvec_S - \xvecNUT| \= 2 |\xvec_N - \xvecNUT|.
\ee
From the $5d$ perspective, which we analyze in more detail below, this critical point can be thought of as the black ring degenerating onto the tip of Taub-NUT, 
since at this point the base space distance between the NUT pole and the middle point between the north and south poles that make up the ring goes to zero. 
The interpretation of the disappearance of this saddle at higher temperatures is unclear to us at this moment, 
and we will speculate about this seeming jump of the index in the discussion in Section \ref{sec:discussion}.

\section{Index saddles for black holes, rings, and lenses in 
asymptotically $5d$ flat space}
\label{sec:index-saddles_flatspace}

In this section, we use the singular four-dimensional saddles \eqref{eq:5d_attractor_saddles_harmonic} discussed in section \ref{sec:three_centers_general} to construct smooth asymptotically 
flat saddles of five-dimensional supergravity coupled to an arbitrary number of vector multiplets. 

To obtain $5d$ asymptotically flat saddles from the M-theory uplift, we need to consider a $4d$ saddle with total asymptotic D6 charge $\Gamma^0 = 1$ 
and take the limit in which the M-theory circle with radius $R_\text{M-theory}$ gets decompactified~\cite{Gaiotto:2005gf}.
To ensure that we obtain a finite temperature $5d$ saddle after decompactification, we need to keep the five-dimensional inverse temperature $\beta/\ell_5$ finite as we take the limit~\cite{Boruch:2025qdq}. 
As measured in five dimensions, the M-theory radius is computed in terms of $4d$ quantities as 
\be 
\frac{R_\text{M-theory}}{\ell_5} \= \frac{\ell_5^2}{\ell_4^2} \= 4\pi (2 \tilde V_{\IIA, \infty})^{1/3} 
\,,
\ee
where we recall that $\tilde V_{\IIA, \infty}$ is the asymptotic volume of the Calabi-Yau manifold measured in string units. 
From the perspective of the four-dimensional supergravity theory, this volume is captured by the asymptotic values of vector multiplet scalar fields $t^A|_\infty = b^A + \i j^A$ via
\be 
\tilde V_{\IIA, \infty} \= \frac{1}{6} D_{ABC} j^A j^B j^C \,. 
\ee
The decompactification limit is then taken by scaling the asymptotic scalars and inverse temperature measured in four dimensions $\beta/\ell_4$ 
with a large parameter $\Lambda$ as
\be 
b^A \;\sim\; O(1) \,, 
\qquad 
j^A \;\sim \; \Lambda^2 \,, 
\qquad 
\frac{\beta}{\ell_4} \;\sim \; \Lambda \,, 
\label{eq:decompactification_limit_scaling}
\ee
and then taking $\Lambda \to \infty$. 
In this limit,  the 
distances between the centers scale as 
\be 
|\xvec_i - \xvec_j| \; \sim \; \frac{1}{\Lambda} \,.
\ee
In order to zoom in on the interesting part of the geometry, we rescale the base space coordinates as~$\xvec \to \xvec/\Lambda$.
Similarly, due to scaling of the entropy function $\Sigma(\Lambda H) = \Lambda^2 \Sigma(H)$, a natural rescaling of the time coordinate is given by $t \to \Lambda t$, which is also consistent with the scaling of $4d$ temperature described above. Because of this, it will be convenient below to rename the inverse temperature as $\beta \to \Lambda \beta$ to keep track of the renormalized inverse temperature.

In the end, after going to rescaled coordinates and taking the limit $\Lambda \to \infty$, the decompactified $5d$ solutions take exactly the same form as \eqref{eq:5d_uplift_formula_1}, with the only difference being that the constant $h$ entering the harmonic function $H(\xvec)$ is reduced only to half of its original components \cite{Cheng:2008gx,Cheng:2006yq} $h_A,h_0 \neq 0$,
\be 
h \= \frac{1}{\Lambda \sqrt{\frac{4}{3}j^3}} 
\left( 0 , 0, - \frac{\Gamma^0}{|\Gamma^0|} j_A^2 
, \frac{1}{|\Gamma^0|} \Gamma^A j_A^2  
\right) 
\= \left( 
0,0, h_A , - \frac{\Gamma^A}{\Gamma^0} 
h_A
\right)
,
\ee
with $\Gamma$ denoting the total 
charge of the configuration.

All solutions constructed in this section can be viewed as saddles of the same index for specific choices of total charge $\Gamma$, related to $5d$ charges $(Q_A^{\text{(5d)}},J_L)$ in presence of a flux sourcing $P^A$ through~\cite{Gaiotto:2005gf},\footnote{The charge $ Q_A^{\text{(5d)}}$ is the electric Page charge of the saddle measured at asymptotic infinity; this can be set by fixing the field strength at asymptotic infinity. The charge $P^A$, a magnetic charge in $4d$, can also be fixed in $5d$ by fixing Dirichlet boundary conditions for the metric and fixing the field strength; by writing the induced metric over the $S^3$ part of the boundary metric in Hopf coordinates, we can fix the integral of the field strength over the $S^2$ base space of this Hopf fibration, $\int_{S^2} F = 4\pi P^A$. }
\begin{align}
\Gamma_\text{total} &\= (\Gamma^0 = 1 , P^A , Q_A, Q_0) 
\\
&\= 
\left(1 , P^A , 
Q_A^{\text{(5d)}} + \frac{D_{ABC}P^B P^C}{2} , 
-\frac{D_{ABC}P^A P^B P^C}{6} 
- P^A Q_A^{\text{(5d)}} - 2J_L
\right) \,.
\end{align}

The $5d$ flat space index can then be obtained by summing contributions from all of these saddles as 
\be 
\mathcal{I}_{\mathbb{R}^4 \times S^1} \= \# e^{-I_{\text{BH}}} + \# e^{-I_{\text{BR}}}
+ \# e^{-I_{\text{BL}}}
+ \# e^{-I_{\text{BSat}}}  + \dots 
.
\ee
As mentioned in the introduction, the competition between saddles contributing to the index may, in some instances, seem at odds with comparisons made between Lorentzian solutions. There is, however, no contradiction: each case naturally corresponds to a different ensemble. For the index, the appropriate description is the grand-canonical ensemble with fixed $J_L$ and $\Omega_R$, whereas Lorentzian solutions are compared in the microcanonical ensemble with fixed $J_L$ and $J_R$. In one instance that will be relevant below, the Lorentzian BMPV black hole cannot be directly compared with the supersymmetric black ring or black lens, since the former has $J_R=0$ while the latter has $J_R\neq 0$ \cite{elvang2004supersymmetric,Kunduri:2014kja}. In the index computation, by contrast, this is not a problem. Moreover, in theories with several vector multiplets, different black rings can be compared microcanonically to one another \cite{Elvang:2004ds}, but these comparisons need not bear a relation to those made for their indices.

We begin in subsection~\ref{sec:flat_space:singleBH} with a brief review of black hole saddles carrying a single unit of D6 charge with $S^3$ horizon. 
We then proceed to explore a new class of solutions obtained from general three-center $4d$ configurations constructed above. 
In subsection~\ref{sec:flat_space:blackRing}, we construct a supersymmetric finite temperature large black ring saddle with $S^2 \times S^1$ horizon by considering a $4d$ bound state of a center carrying unit D6 charge and a large black hole with no D6 charge. 
In subsection~\ref{sec:flat_space:blackLense}, we split the total unit D6 charge between an anti-NUT and a black hole with two units of D6 charge. 
This leads to a $5d$ asymptotically flat space saddle with $S^3/\mathbb{Z}_2$ horizon. 
Finally, in subsections~\ref{sec:flat_space:blackSaturn} and~\ref{sec:flat_space:blackLenseAndblackHole} we discuss explicit examples of $5d$ flat space saddles obtained from $4d$ configurations with four centers. 
This allows us to construct a $5d$ index saddle for a black Saturn, a bound state of a black hole with $S^3$ horizon and a large black ring with $S^2 \times S^1$ horizon, as well as a bound state of a black lens with 
$S^3/\mathbb{Z}_2$ horizon and a black hole with $S^3$ horizon.

\subsection{Saddle I: The BMPV black hole saddle}
\label{sec:flat_space:singleBH}

We briefly review the construction of the black hole saddle with $S^3$ horizon \cite{Boruch:2025qdq}.
To construct a $5d$ black hole index saddle, we consider a $4d$ configuration with two centers, a north pole and a south pole, 
\be 
H(x) \= h  + \frac{\gamma_N}{|\xvec - \xvec_{N}|}
+ \frac{\gamma_S}{|\xvec - \xvec_{S}|} \;,
\ee
with a total monopole charge $\GammaBH=\Gamma_\text{total}$.
The coefficients $\gamma_N \equiv \gamma_1$ and $\gamma_S \equiv \tgamma_1$ of the poles in the harmonic function are given in terms of monopole charge through 
the new attractor equations~\eqref{eq:4d_attractor_saddle_charges}. 
The smoothness at the poles is guaranteed by the conditions
\be 
\i \langle \gamma_N , H(\xvec_N) \rangle = \frac{\beta}{4\pi} ,
\qquad 
\i \langle \gamma_S , H(\xvec_S) \rangle = -\frac{\beta}{4\pi} ,
\ee
which guarantee the removability of the Dirac-Misner string of the one-form $\omega_E$ stretching between the north and south poles in the $4d$ geometry \eqref{eq:4d_attractor_saddle_metric}, and which act as constraints on possible positions of the centers. 
In the current case, they constrain the distance between the north and south poles of the decompactified solution to be
\be 
|\xvec_N - \xvec_S| \= \frac{\langle \GammaBH , \deltaBH \rangle}{\frac{\beta}{4\pi} - \langle h, \deltaBH \rangle} \,,
\label{eq:BH_distance}
\ee
where the dipole charge $\deltaBH$ is determined by~\eqref{eq:4d_attractor_saddle_charges}.
As explained in~\cite{Boruch:2025qdq}, 
smoothness in five dimensions at a pole requires quantization of the D6 charge,
$\Gamma^0 \in \mathbb{Z}$, while achieving $\mathbb{R}^4 \times S^1_{\beta}$ asymptotics demands a single unit of total asymptotic D6 charge $\Gamma_{\text{total}} = 1$. 
For a single black hole constructed from two centers, these conditions fix $\Gamma^0 = 1$, leading to a smooth five-dimensional metric given by \eqref{eq:5d_uplift_formula_1}. This describes a black hole background with an $S^3$ horizon and $\mathbb{R}^4 \times S^1_{\beta}$ asymptotics. In later sections, we will consider configurations with more than two centers, allowing for more general distributions of D6 charge among them.

Imposing the positivity of the distance \eqref{eq:BH_distance}, implies that the constructed saddle is only valid for low enough temperatures in the regime
\be 
\beta \; > \; \beta_{\text{BH}}\,, \qquad \beta_{\text{BH}} \; \equiv \;  4\pi \langle h, \deltaBH \rangle \,. 
\label{eq:BlackHole_cric_temperature}
\ee
This is a familiar constraint arising for many gravitational index saddles, originally observed for supersymmetric rotating Kerr-Newman black holes in \cite{Iliesiu:2021are}. 
The contribution of the saddle to the path integral is now weighted by the on-shell action of the Euclidean saddle. 
As explained in~\cite{Boruch:2025qdq}, the on-shell action can be calculated and, remarkably, the result reproduces the 
entropy~\eqref{eq:extremal_entropy_in_Sigma} of the BMPV black hole~\cite{Breckenridge:1996is}, i.e.,
\begin{align}
- I_{\text{on-shell}}^\text{BH}(\GammaBH) &\= \pi \Sigma(\GammaBH) \= \pi \sqrt{\Q_D^3- L_D^2} \;, \label{eq:5d_flat_space_BH_entropy}
\\ 
L_D(\GammaBH)&\= 
\frac{1}{3} D_{ABC} P^A P^B P^C - P^A Q_A
-Q_0
\, , 
\\
\Q_D^{3} &\= \left( \frac{1}{3} D_{ABC} \, y^A y^B y^C 
\right)^2 ,
\\
D_{ABC}\, y^A y^B &\= D_{ABC} P^A P^B - 2 Q_C 
\,.
\end{align}
The resulting configuration carries classical angular momentum along the $\phi$-circle 
\be 
\textbf{J}_R = 
\frac{\i}{2} \langle
h, \deltaBH
\rangle (\xvec_N - \xvec_S) 
,
\label{eq:JR-black-hole-BMPV}
\ee 
which vanishes in the extremal limit $\beta \to \infty$. 
For a specific comparison with the black ring and black lens saddles that we will make in later subsections, we note the entropy in the D6-D4-D0 black hole,
\be 
 S_{\text{BMPV}} \= 
 \pi \sqrt{\frac{2}{3}D_{ABC}P^A P^B P^C Q_0 - (Q_0)^2},
 \label{eq:D6D4D0_BMPV_entropy}
\ee
with the $4d$ charges $(Q_A, Q_0)$ determined from $5d$ charges $(Q_A^{\text{(5d)}}, J_L)$ in the presence of the flux sourcing D4 charge $P^A$ via
\be 
Q_A^{\text{(5d)}} = - \frac{1}{2} D_{ABC} P^B P^C 
, 
\qquad
J_L = -\frac{Q_0}{2} - \frac{D_{ABC}P^A P^B P^C}{12} .
\label{eq:D6D4D0_charge_relation}
\ee
Note that while the D2 charge is zero, there is a non-zero $Q_A^{\text{(5d)}}$ that is cancelled due to D4 fluxes.\footnote{The gauge transformations of footnote \ref{foot:gaugeM} can be used to fully remove the D4 charges of black objects carrying a D6 charge. From this perspective, our BMPV example is related by an M-theory gauge transformation (with $k^A = P^A$) to a D6-D2-D0 black hole.}

\subsection{Saddle II: A black ring}
\label{sec:flat_space:blackRing}

We now begin the investigation of uplifted solutions in which some of the extremal centers with vanishing area do not get split as we turn on finite temperature. 
The simplest starting point is that of a black ring, where the unit of asymptotic D6 charge is carried by a single center, 
while all the other charges are carried by a $4d$ black hole that 
uplifts to a black ring in five dimensions. 
This charge arrangement is
\be 
\GammaNUT \= ( 1,0,0,0 ) \,, 
\qquad 
\GammaBR \= ( 0 ,P^A , Q_A ,Q_0 ) \,, \qquad \text{ with } \quad \Gamma_\text{total} = \GammaNUT+ \GammaBR \,,
\ee
and correspondingly the harmonic function in \eqref{eq:4d_attractor_saddle_metric} for the finite-temperature solution is
\be
H(x) \= h + \frac{\GammaNUT}{|\xvec-\xvecNUT|} + \frac{\gamma_N}{|\xvec - \xvec_{N}|}
+ \frac{\gamma_S}{|\xvec - \xvec_{S}|} \; .
\ee
The single center $\xvecNUT$ lifts to a nut in five dimensions, where, since the charge $\GammaNUT=1$, the M-theory circle smoothly contracts to a point. 
In contrast, the black ring carries no D6 charge and therefore does not fiber nontrivially over the M-theory circle, as shown in~\cite{Boruch:2025qdq}. Consequently, its horizon has topology $S^2 \times S^1$. The total asymptotic charge is
\be 
\Gamma_{\text{total}} \= \GammaNUT + \GammaBR 
\=(1 , P^A , Q_A , Q_0)
\,,
\ee
which leads to~$\mathbb{R}^4 \times S^1_{\beta}$ asymptotics. 
Note that, as long as its possible to arrange the charges such that $\Sigma(\GammaBR) > 0$ and $\Sigma(\Gamma_{\text{total}}) > 0$, the above total charge can be chosen to be the same as the charge for the black hole discussed in the previous subsection. 

The smoothness of this configuration is ensured by requiring the removability of the Dirac-Misner string stretching between the north and south poles~\eqref{eq:5dsmoothness}, as well as the complete absence of Dirac-Misner strings from the unsplit center $\GammaNUT$~\eqref{eq:5dintegrability}.

The smoothness conditions can be exactly solved for three distances describing the configuration, as explained in section \ref{sec:three_centers_general}. The resulting distances in the decompactified geometry are given by 
\begin{align}
|\xvec_N - \xvecNUT| \= |\xvec_S - \xvecNUT| &\= 
- \frac{\langle \GammaBR , \GammaNUT \rangle}{\langle \GammaBR, h \rangle}
\= \frac{Q_0}{P^A h_A}
\;,
\\ 
|\xvec_N - \xvec_S|
&\=  \frac{\langle \GammaBR, \deltaBR \rangle}{\frac{ \beta}{4\pi}- \langle h ,\deltaBR \rangle - \frac{\langle \GammaNUT , \deltaBR \rangle}{\langle \GammaNUT , \GammaBR \rangle}
\langle \GammaBR, h \rangle} 
\;.
\end{align}
This shows the existence of a smooth solution as long as the above lengths can be embedded in $\mathbb{R}^3$ 
as actual distances between three Cartesian points. 
This is achieved by imposing positivity and triangle inequalities. The positivity of the lengths imposes \eqref{eq:three_center_distance_condition} and \eqref{eq:three-center_temperature_condition},
whereas the triangle inequalities impose \eqref{eq:three-center_temperature_condition_2}.
In parallel to the discussion following \eqref{eq:three-center_temperature_condition}, $\beta_{\text{cr},2}>\beta_{\text{cr},1}$ and, therefore, 
it is sufficient to impose~\eqref{eq:three-center_temperature_condition_2} for the existence of solutions.

The on-shell action of the black ring now evaluates to the entropy of the extremal black ring
\be 
- I_{\text{on-shell}}^\text{BR}(\Gamma_\text{total}) \= 
\pi \Sigma (\GammaBR) \= \pi \sqrt{\frac{1}{3}D_{ABC}P^A P^B P^C 
\bigl(\, 2Q_0 + Q_A \, D_\Gamma^{AB} \, Q_B \bigr)
}
\label{eq:5d_flat_space_BlackRing_entropy}
\; ,
\ee
where $D_\Gamma^{AB}$ is the inverse matrix of $D_{ABC}P^C$. 
The black ring carries classical angular momentum along the $\phi$-circle 
\be 
\textbf{J}_R \= 
\frac{1}{2} \langle \GammaNUT , h \rangle \xvecNUT 
+
\frac{1}{4} \langle \GammaBH , h \rangle (\xvec_N + \xvec_S)
+ \frac{\i}{2} \langle
h, \deltaBH
\rangle (\xvec_N - \xvec_S) 
\, .
\label{eq:JR-black-ring}
\ee
Note that in the extremal limit, the black ring carries nonzero overall angular momentum $J_R = \frac{1}{2} |Q_0|$.

At the critical value $ \beta =  \beta_{\text{BR},2}$, the triangle inequalities saturate and the points become colinear. This implies that the distance between the NUT and the finite temperature black hole horizon goes to zero, since
\be 
\abs{\xvecNUT - \frac{\xvec_N + \xvec_S}{2}} \to 0 \,.
\ee
It therefore looks as if the finite temperature supersymmetric black ring degenerates onto the NUT and then stops existing at higher temperatures. 
However, note that while in the Lorentzian setting the degenerate limit of the black ring corresponds to the BMPV black hole, the degenerate limit of the black ring index saddle differs from the BMPV index saddle: in the latter, the nut center is split, whereas in the former it remains unsplit. It is not clear whether this solution could be obtained starting from a Euclidean rotating non-supersymmetric black ring and imposing $\beta \Omega_R=2\pi i$.

In the D6-D4-D0 case, the black ring's saddle on-shell action is given by
\be 
- I_{\text{on-shell}}^\text{BR}(\Gamma_\text{total}) \= \pi \sqrt{\frac{2}{3}D_{ABC}P^A P^B P^C  Q_0}
\; ,
\label{eq:D6D4D0_black_ring_entropy}
\ee
with the $4d$ charges determined through \eqref{eq:D6D4D0_charge_relation}. A comparison between \eqref{eq:D6D4D0_BMPV_entropy} and \eqref{eq:D6D4D0_black_ring_entropy} shows that the BMPV black hole formed out of $4d$ D6-D4-D0 charges is subdominant to the black ring when viewed as saddles contributing to the gravitational index. This is the index enigma that we alluded to in the introduction. The two solutions were not compared before in Lorentzian signature because they have different values of the right-moving angular momenta: $J_R = 0$ for BMPV, and $J_R = \frac{1}{2}|Q_0|$ for the black ring.\footnote{As in the case of the BMPV black hole analyzed above, in M-theory the total charges $P^A$ can be eliminated through a large gauge transformation (footnote~\ref{foot:gaugeM}). Since the black ring itself does not possess D6 charge, this transformation does not change its charges and $P^A$ can still be measured by the magnetic flux on an $S^2$ that encloses the black ring but not the NUT. The transformation modifies the charges at $\xvecNUT$, resulting in a purely fluxed solution there.  }

\subsection{Saddle III: A black lens}
\label{sec:flat_space:blackLense}

Keeping the same total charge as in previous subsections, we can generate further flat space saddles by considering different assignments of total D6 charge. An interesting class of saddles obtained in this way is that of black lenses \cite{Kunduri:2016xbo}, where a large black hole, referred to as a black lens, carries more than a single unit of D6 charge. The simplest example can be formed out of monopole charges 
\be 
\GammaANUT \= ( -1 ,0,0,0 ) \,, 
\qquad 
\Gamma_{\text{BL}} \= ( 
2, P^A , Q_A , Q_0 
) \,, \quad \text{ with }\quad \Gamma_\text{total} = \GammaANUT + \Gamma_{\text{BL}},
\ee
and a three-pole finite-temperature harmonic function 
\be 
H(x) \= h + \frac{\GammaANUT}{|\xvec-\xvecANUT|} + \frac{\gamma_N}{|\xvec - \xvec_{N}|}
+ \frac{\gamma_S}{|\xvec - \xvec_{S}|} \;. 
\ee
Here, the two units of nut charge in $\Gamma_{\text{BL}}$ get split in the index geometry, but the one in $\GammaANUT$ does not. At the unsplit pole $\xvecANUT$, the M-theory circle again contracts to a point, but now the fibration near the NUT is in the opposite direction of the fibration at the black lens. This results in the same asymptotics as for black rings. The smoothness conditions take the same form as in \eqref{eq:5dsmoothness}, \eqref{eq:5dintegrability}, and correspondingly the distances are now fixed to
\begin{align}
|\xvec_N - \xvecANUT| \= |\xvec_S - \xvecANUT| &= 
- \frac{\langle \GammaBL , \GammaANUT \rangle}{\langle \GammaBL, h \rangle}
\= \frac{Q_0}{P^A h_A}
,
\\ 
|\xvec_N - \xvec_S|
&\=  \frac{\langle \GammaBL, \deltaBL \rangle}{\frac{ \beta}{4\pi}- \langle h ,\deltaBL \rangle - \frac{\langle \GammaANUT , \deltaBL \rangle}{\langle \GammaANUT , \GammaBL \rangle}
\langle \GammaBL, h \rangle} 
\;.
\end{align}
As before, the positivity of the distances imposes \eqref{eq:three_center_distance_condition} and triangle inequalities require temperatures low enough to satisfy \eqref{eq:three-center_temperature_condition_2}.
If any of these conditions is violated, the smoothness equations admit no solution, and the saddle is lost. 

When there are valid solutions, the on-shell action evaluates to the extremal entropy of the black lens, given by~\cite{Boruch:2025biv}
\begin{align}
- I_{\text{on-shell}}^\text{BL}(\Gamma_\text{total}) &\= \pi \Sigma(\GammaBL) \= \frac{\pi}{2} \sqrt{\Q_D^3- L_D^2} \;, 
\label{eq:5d_flat_space_BlackLens_entropy}
\\ 
L_D(\GammaBL)&\= 
\frac{1}{3} D_{ABC} P^A P^B P^C - 2 P^A Q_A
-4Q_0
\,, 
\\
\Q_D^{3} &\= \left( \frac{1}{3} D_{ABC} \, y^A y^B y^C 
\right)^2 ,
\\
D_{ABC}\, y^A y^B &\=  D_{ABC} P^A P^B - 4 Q_C 
\,.
\end{align}

To compare with BMPV and black ring saddles above, we now consider the D6-D4-D0 case discussed in the previous subsections. Then, the black lens on-shell action simplifies to 
\be 
- I_{\text{on-shell}}^\text{BL}(\Gamma_\text{total}) = \pi \sqrt{\frac{2}{3}D_{ABC}P^A P^B P^C Q_0 -4 (Q_0)^2} ,
\ee
showing that, in this case, it is subleading to both the contribution of the black ring \eqref{eq:D6D4D0_black_ring_entropy} and the BMPV black hole \eqref{eq:D6D4D0_BMPV_entropy}.

\subsection{Saddle IV: A black Saturn}
\label{sec:flat_space:blackSaturn}
 We now turn to configurations describing bound states of two large black holes in four dimensions, which uplift to black Saturns in five dimensions \cite{Gauntlett:2004wh,Bena:2004de}. 

These index saddles contain four centers  \cite{Boruch:2025biv}: two north poles and two south poles. To ensure asymptotically flat boundary conditions in five dimensions, we take the monopole charges to be 
\be 
\Gamma_1 \= 
\Gamma_{\text{BH}} \= 
( 1 ,P_{\text{BH}}^A ,
Q_{\text{BH},A}
, Q_{\text{BH},0}
)
\,,
\qquad
\Gamma_2 \=
\Gamma_{\text{BR}} \= ( 0 ,P_{\text{BR}}^A ,
Q_{\text{BR},A}
, Q_{\text{BR},0}
)
\,,
\label{eq:BSat_monopole_charges}
\ee
and their harmonic functions are 
\be 
H(x) \= h +\frac{\gamma_1}{|\xvec - \xvec_1|}
+ \frac{\tgamma_1}{|\xvec - \txvec_{1}|}
+\frac{\gamma_2}{|\xvec - \xvec_2|}
+ \frac{\tgamma_2}{|\xvec - \txvec_{2}|}
\;.
\ee
The north and south poles of both black holes are now connected in pairs by Dirac-Misner strings, whose removability is guaranteed by \eqref{eq:5dsmoothness}.

\begin{figure}[h!]
    \centering
    \includegraphics[height=7.5cm]{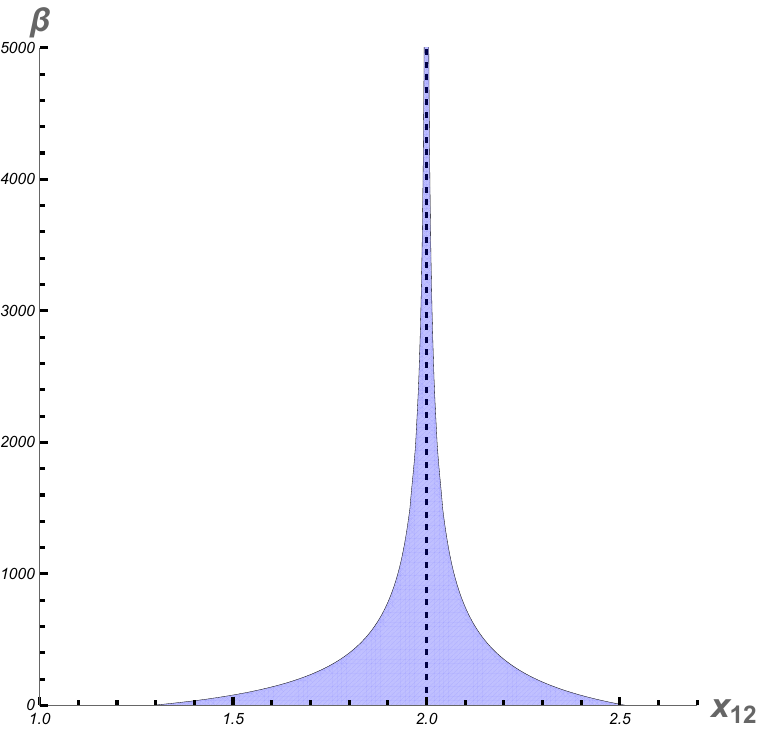}
    \includegraphics[height=7.5cm]{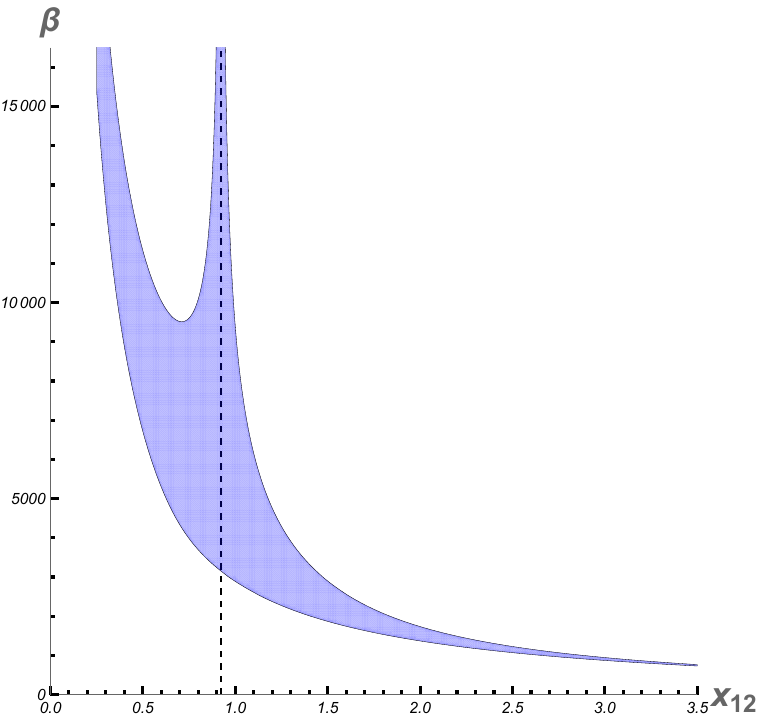}
    \caption{Examples of allowable values of $x_{12}$ distance as a function of $\beta$ for the case of black Saturn in asymptotically flat $5d$ space, as determined by the smoothness of the saddle, for two different choices of total charges. In both figures, the dashed lines represent the distance which matches the extremal bound state distance $x_{12}^{*} = -\langle \Gamma_1 , \Gamma_2 \rangle/\langle \Gamma_1, h \rangle$. \textbf{Left:} For the choice of charges $\Gamma_1 = (1,9,3,3) $, $\Gamma_2 = (0,6,3,3)$, as we lower the temperature, the space of solutions shrinks until it collapses on the extremal value of the distance. \textbf{Right:} For the choice of charges $\Gamma_1 = (1,2,1,20) $, $\Gamma_2 = (0,13,10,5) $, 
    as we lower the temperature, the moduli space becomes disconnected, leading to two possible extremal solutions. The solution corresponding to the left peak is a scaling solution where all base-space distances shrink as $\sim 1/\beta$. }
    \label{fig:BlackSaturn}
\end{figure}

The moduli space defined by these equations was studied in detail in \cite{Boruch:2025biv}. For four centers, there are six independent inter-center distances, precisely matching the number of coordinates needed to specify the configuration of four points once overall translations and rotations are fixed. One can therefore solve the equations directly for the distances and, after checking that they can be embedded in $\mathbb{R}^3$, solve for the coordinates of the north and south poles. The explicit solutions for the distances are given by 
\begin{align}
|\xvec_1 - \txvec_2| &\= 
\frac{B_{1 \twobar}}{d_1 - \frac{B_{12}}{|\xvec_1 - \xvec_2|}} \,,
\qquad 
|\txvec_1 - \xvec_2| \= 
|\xvec_1 - \txvec_2|
\,,
\qquad 
|\txvec_1 - \txvec_2| \= |\xvec_1 - \xvec_2|\,, 
\label{eq:solutions_regularity_conditions1}
\\ 
|\xvec_1 - \txvec_1| &\=  
\frac{A_{1\onebar} B_{1\twobar}}{(B_{1\twobar}c_1-A_{1\twobar}d_1)
+ \frac{A_{1\twobar}B_{12}+A_{12}B_{1\twobar}}{|\xvec_1 - \xvec_2|}}
\,, 
\label{eq:solutions_regularity_conditions2}
\\ 
|\xvec_2 - \txvec_2| &\= 
\frac{A_{2\twobar} B_{1\twobar}}{(B_{1\twobar}c_2-A_{1\twobar}d_1)
+ \frac{A_{1\twobar}B_{12}-A_{12}B_{1\twobar}}{|\xvec_1 - \xvec_2|}}
\,,
\label{eq:solutions_regularity_conditions3}
\end{align}
where we denoted 
\begin{align}
B_{1\twobar}&\= 
\frac{\langle \Gamma_1,\Gamma_2 \rangle}{4}+
\langle \delta_1,\delta_2 \rangle 
\,, 
\qquad
B_{12}\=
\frac{\langle \Gamma_1,\Gamma_2 \rangle}{4}-
\langle \delta_1,\delta_2 \rangle  
\,, 
\qquad
A_{1\onebar} = 
\langle \Gamma_1 , \delta_1 \rangle ,
\\
A_{1 \twobar}&\= 
\frac{\langle \Gamma_1,\delta_2 \rangle-\langle \delta_1,\Gamma_2 \rangle}{2} 
\,, 
\qquad
A_{1 2}\= 
\frac{\langle \Gamma_1,\delta_2 \rangle+\langle \delta_1,\Gamma_2 \rangle}{2}
\,, 
\qquad 
A_{2\twobar} \= 
\langle \Gamma_2 , \delta_2 \rangle \,,
\\ 
c_1 &\= 
\frac{\beta}{4\pi} - \langle h , \delta_1 \rangle  \,,
\qquad
c_2\= 
\frac{\beta}{4\pi} -  \langle h , \delta_2 \rangle   \,, 
\\
d_1  &\= 
\frac{-\langle \Gamma_1, h \rangle}{2}
\= 
\frac{1}{2} P_{\text{BS}}^A h_A 
\,, \qquad 
d_2 \= 
- d_1
\,.
\label{eq:defd1d2P}
\end{align}
and kept the remaining distance $|\xvec_1 - \xvec_2|$ as the free parameter characterising the continuous solution space. 
By imposing that the distances can be embedded in $\mathbb{R}^3$, the analogue of triangle inequalities, constrains the values of $|\xvec_1 - \xvec_2|$ to a finite temperature-dependent region on a real line. The moduli space can then be viewed as two arrows at finite distance from each other, restricted to point along parallel planes, with all the distances in the configuration fixed to some specific value. 
To verify that there are indeed solutions for the specific case of charges \eqref{eq:BSat_monopole_charges}, we numerically plot the Cayley-Menger determinants for $\Gamma_1 = (1,9,3,3) $, $\Gamma_2 = (0,6,3,3)$, in Figure \ref{fig:BlackSaturn}. This verifies that there exists a regime of the free parameter $x_{12}$ for which the distances \eqref{eq:solutions_regularity_conditions1}, \eqref{eq:solutions_regularity_conditions2}, \eqref{eq:solutions_regularity_conditions3} can be embedded in $\mathbb{R}^3$.

The structure of critical temperatures for which the black Saturn exists is now more complicated, since one wants to ensure the positivity of all of the distances, triangle inequalities for all triangles, and positivity of the Cayley-Menger determinant (plotted in the figure). 
What is clear, though, is that one can always choose low enough temperatures such that all of those conditions are satisfied, and it is also possible to choose a high enough temperature where the saddle disappears. 
The on-shell action of the bound states evaluates now to the sum of black hole and black ring entropies \cite{Boruch:2025biv}
\be 
- I_{\text{on-shell}}^\text{BS}(\Gamma_\text{total}) = \pi \Sigma_{\text{BS}}(\Gamma_1 + \Gamma_2) 
\= 
\pi \Sigma(\GammaBH) \; + \; \pi \Sigma(\Gamma_{\text{BR}}), 
\ee
with explicit expressions given by  \eqref{eq:5d_flat_space_BH_entropy} and \eqref{eq:5d_flat_space_BlackRing_entropy}.

Lastly, we comment on wall-crossing when tuning the asymptotic boundary conditions for the scalar fields. 
In four dimensions, the full moduli space of solutions is lost when $\langle \Gamma_1, h \rangle$ changes sign. 
In the current case, after the decompactification limit, we have that 
\be 
\langle \Gamma_1, h \rangle \= - P_{\text{BR}}^A h_A  ,
\ee
and therefore, in the case when $P_{\text{BR}}^A > 0$, wall-crossing in the scalar moduli cannot occur for the black Saturn in five-dimensional flat space.

\subsection{Saddle V: A black lens and a black hole}
\label{sec:flat_space:blackLenseAndblackHole}

Lastly, we can also consider a bound state of a black lens and a black hole in five dimensions. This corresponds to monopole charges 
\be 
\Gamma_1 \= 
\GammaBH  \= ( -1, P_1^A ,Q_{1,A} ,Q_{1,0} ) 
, \qquad
\Gamma_2 \= 
\GammaBL \= (2 , P_2^A ,Q_{2,A} ,Q_{2,0} ) 
,
\ee
and a four-center harmonic function
\be 
H(x) \= h +\frac{\gamma_1}{|\xvec - \xvec_1|}
+ \frac{\tgamma_1}{|\xvec - \txvec_{1}|}
+\frac{\gamma_2}{|\xvec - \xvec_2|}
+ \frac{\tgamma_2}{|\xvec - \txvec_{2}|}
\,.
\ee
The smoothness conditions are the same as for the black Saturn, and they can be solved similarly. The moduli space will follow from the solutions for the distances \eqref{eq:solutions_regularity_conditions1}-\eqref{eq:solutions_regularity_conditions3}, where we note that the constant $d_1$ is given by
\be 
d_1 \= - \frac{\langle \Gamma_1,h \rangle}{2} 
\= - (P_1^A + \frac{1}{2} P_2^A ) h_A \,.
\ee
This condition shows again that when working with positive D4 charges $P_1^A , P_2^A > 0$ wall-crossing does not occur for the bound state of a black hole and a black lens.

We can numerically verify the existence of a region in $x_{12}$ space where the Cayley-Menger determinant is positive for specific values of the monopole charges, in particular for $\Gamma_1 = (-1,6,3,3) $ and $\Gamma_2 = (2,9,3,3)$, which match the total asymptotic values of the previous subsections.
This shows that for low enough temperatures, both black Saturns and bound states of black holes with black lenses will contribute to the same index. For the charges chosen above, we find that the bound state of black hole and black lens dominates over black saturn,  although there is no general statement about which one will make the larger contribution. For high temperatures, one will again encounter a critical temperature beyond which the saddle is lost. 
The on-shell action weighing saddle's contribution to the path integral now evaluates to a sum of black lens and black hole entropy \cite{Boruch:2025biv}
\be 
- I_{\text{on-shell}}^\text{BH+BL}(\Gamma_\text{total}) = \pi \Sigma_\text{BH+BL}(\Gamma_1 + \Gamma_2) 
\= \pi \Sigma(\GammaBH)
\; + \; \pi \Sigma(\GammaBL)
\,.
\ee
We note that for the negative D6-charge, the black hole entropy is given by
\begin{align}
\Sigma(\GammaBH) &\= \sqrt{\Q_D^3- L_D^2} \;, 
\label{eq:5d_flat_space_BH_entropy_OPPOSITE}
\\ 
L_D(\GammaBH)&\= 
\frac{1}{3} D_{ABC} P^A_1 P^B_1 P^C_1 + P^A_1 Q_{1,A}
-Q_{1,0}
\;, 
\\
\Q_D^{3} &\= \left( \frac{1}{3} D_{ABC} \, y^A y^B y^C 
\right)^2 ,
\\
D_{ABC}\, y^A y^B &\= D_{ABC} P^A_1 P^B_1 + 2 Q_{1,C} 
\,.
\end{align}

\section{Index saddles with AdS$_3 \times S^2$ asymptotics}
\label{sec:indices-in-ads3-s2}

We now turn to the construction of five-dimensional index saddles with AdS$_3 \times S^2$ asymptotics. Since this requires that there is no Hopf fibration over the $S^2$, the D6 charge of the four-dimensional configurations must vanish, $\Gamma^0 = 0$. To obtain these saddles, we first perform the M-theory uplift and then take the decoupling limit of the resulting configuration. This method was employed for extremal solutions in \cite{deBoer:2008fk,VandenBleeken:2008tsa}, and it will now give us the corresponding finite-temperature index saddles. 

To take the decoupling limit, we first scale the asymptotic $4d$ boundary conditions as
\be 
b^A \sim O(1) , 
\qquad 
j^A \sim \Lambda^2 , 
\qquad 
\frac{\beta}{\ell_4} \sim \Lambda^3 , 
\ee
and then take $\Lambda \to \infty$. 
This is 
similar to the scaling~\eqref{eq:decompactification_limit_scaling} discussed in the previous section. However, now there is a different scaling for the inverse temperature $\beta/ \ell_4$. 
This new scaling follows from the fact that when the total D6 charge is zero, the distances between poles at large $\Lambda$ behave as 
\be 
|\xvec_i - \xvec_j| \;\sim\; \frac{1}{\Lambda^3}.
\ee 
Similar to the previous section, if we want the limit to zoom in on the interesting part of the geometry
we must rescale the base space coordinates $\xvec \to \xvec/\Lambda^3$ and $t \to \Lambda^3 t$.
The latter implies that the rescaled inverse temperature is obtained as $\beta \to \Lambda^3 \beta$. 

The resulting geometry has the form~\eqref{eq:5d_uplift_formula_1} and  \eqref{eq:4d_attractor_saddle_metric}, 
with the 
constant in the harmonic function given by
\be \label{eq:hdec}
h \= 
\frac{1}{\Lambda^3}
\left( 0 , 0, 0
, 
\frac{\sqrt{j^3}}{2\sqrt{3}}
\right) 
\= \left( 0,0,0,h_0 \right) 
\,.
\ee

In Section~\ref{sec:AdS3_Asymptotics}, we first show that, in the decoupling limit, a general multicentered configuration with vanishing total D6 charge exhibits AdS$_3 \times S^2$ asymptotics.
With appropriate choices of charges, such configurations yield a large family of smooth supersymmetric index saddles. In Section~\ref{sec:BTZS2} we briefly review the simplest solution with this asymptotics: the BTZ$\times S^2$ saddle.
We then analyze several explicit and illustrative examples of these new solutions.
In Section~\ref{sec:S^3_in_AdS3}, we study a BMPV black hole with an $S^3$ horizon inside AdS$_3 \times S^2$, while in Section~\ref{sec:blackLens_in_AdS3} we construct a black lens in AdS$_3 \times S^2$ by combining it with two anti-NUT centers.
Finally, in Section~\ref{sec:twoBMPV_in_AdS3}, we discuss the bound state of two black holes with opposite D6 charges and horizons of topology $S^3/|\mathbb{Z}_{P^0}|$.

\subsection{Asymptotics of general multicentered configuration}
\label{sec:AdS3_Asymptotics}

In this section, we show that a multicentered configuration with a vanishing total D6 charge develops AdS$_3\times S^2$ asymptotics in the decoupling limit described above. 
The analysis follows closely the zero temperature case originally derived in \cite{deBoer:2008fk}.
For a generic configuration, the harmonic functions at infinity behave as
\begin{align}
H^0 &\=  \frac{\mathbf{d}^0 \cdot \mathbf{e}}{r^2} + O(r^{-3}) 
= \frac{d^0 \cos \theta}{r^2} + O(r^{-3}) 
\,, 
\\ 
H^A &\= \frac{\Gamma^A}{r} + \frac{\mathbf{d}^A \cdot \mathbf{e}}{r^2} + O(r^{-3})
\,, 
\\ 
H_A &\= 
\frac{\Gamma_A}{r} + \frac{\mathbf{d}_A \cdot \mathbf{e}}{r^2} + O(r^{-3})
\,, 
\\ 
H_0 &\=  h_0 + \frac{\Gamma_0}{r} + \frac{\mathbf{d}_0 \cdot \mathbf{e}}{r^2} + O(r^{-3})
\,,
\end{align}
where we denoted the dipole moments as 
\begin{align}
\mathbf{d}^\alpha
&\; \equiv \; 
\sum_{i=1}^{N_{NUT}} \Gamma_{\text{NUT},i}^\alpha \xvec_{\text{NUT},i}
+
\sum_{i=1}^{N_{BH}} (\gamma_i^\alpha \xvec_i 
+ \tgamma_i^\alpha \txvec_i) \,,
\end{align}
$\mathbf{e} \equiv \{\frac{x^i}{r}\}$, and in the first line we have aligned the $z$-axis with the dipole moment $\mathbf{d}^0$. This is a natural choice as the angular momentum ``points" in the direction of $\mathbf{d}^0$
\begin{align}
\mathbf{J}_R &\= \frac{1}{2} \sum_{i=1}^{N_{NUT}} \langle h, \Gamma_{\text{NUT},i} \rangle 
\xvec_{\text{NUT},i}
+
\frac{1}{2} \sum_{i=1}^{N_{BH}} (\langle h ,\gamma_i \rangle \xvec_i + \langle h ,\tgamma_i \rangle \txvec_i )
\= -\frac{1}{2} h_0 \mathbf{d}^0 \,.
\end{align}
To study the asymptotics of the metric, we need to determine the asymptotic behaviors of the functions $\omega,L_D,\mathcal{A}_d^0$ and $\Q_D$ entering the metric. Using the above form of the harmonic functions, we can easily determine, from \eqref{eq:4d_attractor_saddle_metric}, \eqref{eq:explicit_gauge_field_A0},
\begin{align}
\omega_E &\= \frac{2\i J_R \sin^2 \theta \dd \phi}{r} + O(r^{-2}) 
\= - \frac{\i h_0 d^0 \sin^2 \theta \dd \phi}{r} + O(r^{-2}) 
\,,
\\
\mathcal{A}_d^0 &\= - \frac{d^0 \sin^2 \theta \dd \phi}{r} + O(r^{-2}) \,.
\end{align}
To find the behavior of $\Q_D$, defined through \eqref{eq:Q_function_Shmakova}, we need to solve for the functions $y^A(H)$ near infinity. These are determined from a set of quadratic equations \eqref{eq:y_function_Shmakova} 
which can be solved in perturbation theory as 
\be 
y^A_{\text{pert}} \= 
H^A 
- 
H^0 D^{AD} H_D 
- \frac{1}{2} (H^0)^{2} D^{FA} 
D_{FBC} D^{BD} H_D D^{CE} H_E 
+ O\left(\frac{1}{r^4}\right) 
\,,
\ee
and where we defined $D^{AB}$ as the inverse of the matrix $D_{ABC}H^C$.
One can explicitly verify that, with the harmonic functions taken as above, the following equations are satisfied, 
\be 
D_{ABC} y^A_{\text{pert}} y^B_{\text{pert}} 
\= 
D_{ABC}H^A H^B - 2H_C H^0  + O\left( \frac{1}{r^4} \right) 
.
\ee
Then, the asymptotic behaviors of $L_D$ and $\Q_D$ are given by
\begin{align}
\label{eq:QDLDH}
\Q_D^{3/2} &\= \frac{1}{3} D_{ABC} H^A H^B H^C - 
H^0 H^A H_A + \frac{1}{2} (H^0)^2 H_A D^{AB} H_B + O(r^{-6}) \,, 
\\ 
L_D &\= \frac{1}{3} D_{ABC} H^A H^B H^C - H^0 H^A H_A - (H^0)^2 H_0 \,.
\end{align}
We express the above expansions as
\begin{align}
\Q_D^{3/2} &\= \frac{q_3}{r^3} + \frac{q_4}{r^4} + \frac{q_5}{r^5} + O(r^{-6})
\,,\\
L_D &\= \frac{q_3}{r^3} + \frac{q_4- h_0 d_0^2 \cos^2 \theta}{r^4} + \frac{l_5}{r^5} + O(r^{-6})
\,,
\end{align}
where $q_3$,$q_4$,$q_5$ are constants that can be read off from~\eqref{eq:QDLDH}. 
Performing a change of coordinates 
\be 
t \= h_0 \tau 
\,,
\qquad
\psi \= \frac{1}{2}(\sigma - \i \tau ) \,, 
\qquad
r \= \frac{2 U}{h_0} e^{\frac{\eta}{U}} 
- \frac{d^0 \Gamma^A \Gamma_A \cos\theta
- D_{ABC} \Gamma^A \Gamma^B \mathbf{d}^C \cdot \mathbf{e}}{3 U^3} \,,
\ee
and plugging in the above expansions, we find that as $\eta \to \infty$ the metric behaves as 
\begin{align}
\dd s^2_5 &\= \dd \eta^2 + e^{\frac{\eta}{U}} 
(\dd \sigma^2 + \dd \tau^2) + \frac{1}{4U^4}
\left( 
\mathcal{D} (\dd \sigma - \i \dd \tau)^2 
- (h_0 d^0)^2 (\dd \sigma + \i \dd t)^2 
\right) 
\\ 
&\quad +U^2 \left( 
\dd \theta^2 +\sin^2 \theta 
\left( 
\dd \phi- \frac{h_0 d^0}{2 U^3}(\dd \sigma + \i \dd \tau)
\right)^2
\right)
,
\end{align}
where we introduced
\begin{align}
U^3 \= \frac{1}{3} D_{ABC} \Gamma^A \Gamma^B \Gamma^C , 
\qquad 
\mathcal{D} \= \left(\frac{1}{3} D_{ABC} \Gamma^A \Gamma^B \Gamma^C) 
(2\Gamma_0 + \Gamma_A D^{AB}_\Gamma \Gamma_B\right) 
\,,
\end{align}
and $D^{AB}_\Gamma$ denotes the inverse of the matrix $D_{ABC}\Gamma^C$.
This shows that the uplifted metric in the decoupling limit is indeed asymptotically AdS$_3 \times S^2$ 
with a nontrivial fibration over the~$S^2$. 
We now proceed to analyze some of the simplest cases that can be obtained through this construction.

\subsection{Saddle I: BTZ $\times S^2$ in AdS$_3 \times S^2$}\label{sec:BTZS2}

Before discussing more intricate saddles, let us recall the simplest one contributing to the index of the MSW CFT \cite{Maldacena:1997de}, which was described in~\cite{Boruch:2025qdq} (Sec.~5). This configuration is a five-dimensional black string with horizon topology $S^1\times S^2$, carrying monopole charges
\begin{align}
    \Gamma = (0,P^A, Q_A, Q_0)\,.
\end{align}
In the decoupling limit to AdS$_3 \times S^2$ asymptotics, the finite-temperature harmonic functions take the form
\begin{align}
    H(x)=h + \frac{\gamma_N}{|\xvec - \xvec_{N}|}
+ \frac{\gamma_S}{|\xvec - \xvec_{S}|} \,,
\end{align}
with $h=(0,0,0,h_0)$, and with $\gamma_{N,S}$ subject to the standard regularity conditions~\eqref{eq:multicentered_regularity_condition}.

For general charges, the on-shell action evaluates to 
\be 
- I_{\text{on-shell}}^\text{BTZ}(\Gamma_\text{total}) = 
\pi \Sigma (\Gamma_{\text{total}}) \= \pi \sqrt{\frac{1}{3}D_{ABC}P^A P^B P^C 
\bigl(\, 2Q_0 + Q_A \, D_\Gamma^{AB} \, Q_B \bigr)
}
\label{eq:BTZ_on-shell_action_GENERAL}
\; ,
\ee
where $D_\Gamma^{AB}$ is the inverse matrix of $D_{ABC}P^C$. 
In the simplest case with D4 and D0 charges and no D2 charge, $Q_A=0$, Ref.~\cite{Boruch:2025qdq} explicitly showed that the corresponding metric reduces to a rotating BTZ black hole times $S^2$, with action
\be \label{eq:BTZaction}
- I_{\text{on-shell}}^\text{BTZ}(\Gamma_\text{total}) \=
\pi \Sigma_{\text{BTZ}}(\Gamma_\text{total}) \= \pi \sqrt{\frac{2}{3}D_{ABC}P^A P^B P^C Q_0} \; .
\ee

\subsection{Saddle II: $S^3$ horizon in AdS$_3 \times S^2$}
\label{sec:S^3_in_AdS3}
As a first example which showcases the types of solutions that follow from the construction of section \ref{sec:three_centers_general}, we consider a bound state of a black hole with an $S^3$ horizon and a NUT with monopole charges
\begin{align}
\GammaNUT \= \Gamma_1 \= ( 1,0,0,0 ) \,, 
\qquad 
\GammaBH \= \Gamma_2 \= ( -1 , P^A , Q_A , Q_0 ) \, .
\label{bmpvbub}
\end{align}
In the extremal solution, as one approaches the black hole center, the asymptotic AdS$_3 \times S^2$ geometry transitions into AdS$_2 \times S^3$ near the horizon, with the angular direction of AdS$_3$ becoming the Hopf fiber over the $S^2$ at the black hole horizon. The latter is a feature shared by the finite-temperature index saddles.

The finite-temperature harmonic functions are now given by
\be 
H(x) \= h + \frac{\GammaNUT}{|\xvec-\xvecNUT|} + \frac{\gamma_N}{|\xvec - \xvec_{N}|}
+ \frac{\gamma_S}{|\xvec - \xvec_{S}|} \, ,
\ee
and we impose the mixed regularity conditions \eqref{eq:5dsmoothness}, \eqref{eq:5dintegrability}. These fix the base space distances between the poles to 
\begin{align}
|\xvec_N - \xvecNUT| \= |\xvec_S - \xvecNUT| &\= 
- \frac{\langle \GammaBH , \GammaNUT \rangle}{\langle \GammaBH, h \rangle}
\= - \frac{Q_0}{h_0}
\;,
\\ 
|\xvec_N - \xvec_S|
& \=  \frac{\langle \GammaBH, \deltaBH \rangle}{\frac{ \beta}{4\pi}- \langle h ,\deltaBH \rangle - \frac{\langle \GammaNUT , \deltaBH \rangle}{\langle \GammaNUT , \GammaBH \rangle}
\langle \GammaBH, h \rangle} \;.
\end{align}
The positivity of the distances and triangle inequalities again impose \eqref{eq:three-center_temperature_condition_2} and \eqref{eq:three_center_distance_condition}.
The black hole will therefore exist as a saddle of the index only for low enough temperatures below the critical values. 
The contribution to the index will be weighted by the on-shell action of the black hole saddle, which evaluates to extremal entropy \cite{Boruch:2025qdq} determined through \eqref{eq:5d_flat_space_BH_entropy_OPPOSITE}.

The solution is essentially a BMPV black hole inside an AdS$_3 \times S^2$ throat. 
Since the configuration has AdS$_3$ asymptotics, it is described by the dual MSW CFT with central charge given by the D4-brane charge. 
In this configuration, the D4 charge arises entirely from flux on the D6 worldvolume, while the entropy is carried by a BMPV black hole at the tip of the nut, built from $Q_A^{\text{(5d)}}$ D2-branes and $Q_0^{\text{(5d)}}=2J_L$ D0-branes. As we explained in Sec.~\ref{sec:4d-to-5d-smoothness}, the D4 flux on the D6 induces further D2 and D0 charges, contributing to the total $Q_A$ and $Q_0$, so that 
\be 
Q_A\=Q_A^{\text{(5d)}}+\frac12 D_{ABC}P^B P^C \, ,
\qquad
Q_0 \= \frac{1}{3}D_{ABC}P^A P^B P^C + P^A Q_A^{\text{(5d)}} -  Q_0^{\text{(5d)}} \, .
\ee

The purely fluxed charges do not contribute to the entropy, which therefore takes the BMPV form in terms of the $5d$ charges, 
\be\label{bmpvS}
S_{\text{BMPV}} \= \pi \Sigma(\GammaBH) \= \pi \sqrt{\Q_D^3 - 4 J_L^2}\,,
\qquad 
\Q_D^3 = \frac{1}{3}D_{ABC}P^A P^B P^C \,.
\ee
Here it is understood that, although $\Q_D$ can be written in terms of the $4d$ charges of the MSW theory, it is determined by the $5d$ charges through 
\be 
Q_A^{\text{(5d)}} = - \frac{1}{2} D_{ABC} P^B P^C \,.
\ee

For illustration, we set $Q_A=0$, so that there are only D4 and D0 charges. 
Then the on-shell action \eqref{bmpvS} can be written as 
\be 
- I_{\text{on-shell}}^\text{BMPV}(\Gamma_\text{total}) = S_{\text{BMPV}} \= 
\pi \Sigma (\GammaBH) \= \pi \sqrt{\frac{2}{3}D_{ABC}P^A P^B P^C Q_0 - (Q_0)^2} \;.
\ee
This is less than the on-shell action of the BTZ black hole with the same total charges, $\Gamma_\text{BTZ}=( 0,P^A,0,Q_0)$, namely, \eqref{eq:BTZaction},
showing that the configuration \eqref{bmpvbub} with $Q_A=0$ gives a subleading contribution to the MSW index. 

This parallels our discussion in section \ref{sec:index-saddles_flatspace}, where we found that the black ring dominates over the black hole in flat space---the index enigma, which here acquires a more ordinary reading. The BTZ saddle has the same charges as the black ring, while the BMPV saddle has the same charges as the black hole---in both cases, the NUT charge does not affect the value of the on-shell action, its only role being to provide the required asymptotics. The fact that the black ring dominates with this choice of charges in asymptotically flat space implies that BTZ dominates in AdS$_3\times S^2$.  This feature, however, need not generally hold for other values of the D2 charge. Indeed, by tuning the value of the D2 charge, the BMPV black hole can be made to dominate over BTZ.\footnote{To see this, scale the D2 charges $Q_A \to \lambda Q_A$ for large $\lambda$. Then, the on-shell action of BTZ, \eqref{eq:BTZ_on-shell_action_GENERAL}, scales as $\sim \lambda$, while the action of BMPV, \eqref{eq:5d_flat_space_BH_entropy_OPPOSITE}, scales as $\sim \lambda^{3/2}$. Then, for large enough $\lambda$ the BMPV black hole will dominate over the BTZ$\times S^2$ saddle.}

\subsection{Saddle III: a black lens in AdS$_3 \times S^2$}
\label{sec:blackLens_in_AdS3}
The next example we discuss is that of a black lens inside of the AdS$_3 \times S^2$ throat. For this, we consider a black hole with $S^3/\mathbb{Z}_2$ horizon, superposed with two anti-NUT centers. This corresponds to three monopole charges of the type 
\begin{align}
 \Gamma_1 \=  \Gamma_2 \=  \GammaNUT  = (-1,0,0,0 ) \, , 
\qquad
 \Gamma_3 \= \GammaBH \= (2, P^A , Q_A , Q_0 ) \,,
\end{align}
The harmonic functions are given by
\be 
H(x) \= h + \frac{\Gamma_1}{|\xvec-\xvec_{\nut,1}|}
+\frac{\Gamma_2}{|\xvec-\xvec_{\nut,2}|}
+ \frac{\gamma_3}{|\xvec - \xvec_{3}|}
+ \frac{\tgamma_3}{|\xvec - \txvec_{3}|}\; ,
\ee
and the condition for smoothness of the geometry is 
\begin{align} \label{eq:attrBLens}
i\=1,2: 
\qquad
\langle \Gamma_i , H(\xvec_i) \rangle \= 0\,,
\qquad
\i \langle \gamma_3 , H(\xvec_3) \rangle \= \frac{\beta}{4\pi} \,,
\qquad 
\i \langle \tgamma_3 , H(\txvec_3) \rangle \= -\frac{\beta}{4\pi} \,.
\end{align}
There are 6 distances and only 6 parameters that describe the system after fixing rotations and translations, which allows us to first solve the equations for the distances and then determine the Cartesian coordinates of the points.
The system of equations~\eqref{eq:attrBLens} consists of 
three complex, i.e.~six real equations. One of them, however, linearly depends on the others. 
We can thus solve for five out of six distances. 
Below, we chose to keep the distance between anti-NUTs~$x_{12}$ as a free parameter. 
The other distances are determined to be
\begin{align}
x_{13} &\= \frac{\langle \Gamma_3 , \Gamma_1 \rangle}{\langle \Gamma_1 , h \rangle}
\= \frac{Q_0}{-h_0}
\= x_{1\threebar} \,, \qquad
x_{23} \= 
\frac{\langle \Gamma_3 , \Gamma_2 \rangle}{\langle \Gamma_2 , h \rangle}
= \frac{Q_0}{-h_0}
\= x_{2\threebar} \;,
\\
x_{3\threebar} &\= \frac{\langle \Gamma_3 , \delta_3 \rangle}{\frac{\beta}{4\pi}-\langle h, \delta_3 \rangle 
+ \langle \Gamma_1 ,h \rangle \frac{\langle \delta_3 , \Gamma_1 \rangle}{\langle \Gamma_3 , \Gamma_1 \rangle}
+ 
\langle \Gamma_2 ,h \rangle \frac{\langle \delta_3 , \Gamma_2 \rangle}{\langle \Gamma_3 , \Gamma_2 \rangle}
}
\=
\frac{\langle \Gamma_3 , \delta_3 \rangle}{\frac{\beta}{4\pi}+ h_0 \delta_3^0  
- 2h_0 \frac{\delta_{3,0}}{Q_0}} \;,
\end{align}
Note that the expressions for the above five distances are independent of~$x_{12}$ -- this happens because the two NUT charges do not interact with each other $\langle \Gamma_1 , \Gamma_2 \rangle = 0$. 
The range of~$x_{12}$ is only constrained by the Cayley-Menger condition \cite{Boruch:2025biv} which ensures that the above distances can be consistently embedded in $\mathbb{R}^3$. 
As before, the saddle is only valid in a specific regime of low enough temperatures, with specific critical temperatures determined through positivity, triangle inequalities, and the Cayley-Menger condition. 
The on-shell action of allowed solutions precisely equals the extremal entropy of the black lens given by~\eqref{eq:5d_flat_space_BlackLens_entropy}.

\subsection{Saddle IV: two BMPV black holes or black lenses in $AdS_3 \times S^2$}
\label{sec:twoBMPV_in_AdS3}
Finally, we consider a bound state of two BMPV black holes with $S^3/\mathbb{Z}_{|P^0|}$ horizons and opposite values of D6 charges
\begin{align}
 \Gamma_1 \= (P^0,P^A_1 , Q_{1,A} , Q_{1,0} ) \,, 
\qquad
 \Gamma_2  \= (-P^0, P^A_2 , Q_{2,A} , Q_{2,0})\,,
\end{align}
which corresponds to the following harmonic functions
\be 
H(x) \= h
+ \frac{\gamma_1}{|\xvec - \xvec_{1}|}
+ \frac{\tgamma_1}{|\xvec - \txvec_{1}|} 
+
\frac{\gamma_2}{|\xvec - \xvec_{2}|}
+ \frac{\tgamma_2}{|\xvec - \txvec_{2}|} \;.
\ee
Following the same strategy as in four dimensions \cite{Boruch:2025biv} for two north poles and two south poles, the regularity condition can be explicitly solved for five out of six distances, with the remaining distance $|\xvec_1 - \xvec_2|$ chosen as the free parameter on the space of solutions. The explicit distances take the form
\begin{align}
|\xvec_1 - \txvec_2| &\= 
\frac{B_{1 \twobar}}{d_1 - \frac{B_{12}}{|\xvec_1 - \xvec_2|}} \,,
\qquad 
|\txvec_1 - \xvec_2| \= 
|\xvec_1 - \txvec_2|
\,,
\qquad 
|\txvec_1 - \txvec_2| \= |\xvec_1 - \xvec_2|\,, 
\label{eq:solutions_regularity_conditions1P}
\\ 
|\xvec_1 - \txvec_1| &\=  
\frac{A_{1\onebar} B_{1\twobar}}{(B_{1\twobar}c_1-A_{1\twobar}d_1)
+ \frac{A_{1\twobar}B_{12}+A_{12}B_{1\twobar}}{|\xvec_1 - \xvec_2|}}
\,, 
\label{eq:solutions_regularity_conditions2P}
\\ 
|\xvec_2 - \txvec_2| &\= 
\frac{A_{2\twobar} B_{1\twobar}}{(B_{1\twobar}c_2-A_{1\twobar}d_1)
+ \frac{A_{1\twobar}B_{12}-A_{12}B_{1\twobar}}{|\xvec_1 - \xvec_2|}}
\,,
\label{eq:solutions_regularity_conditions3P}
\end{align}
where, as before, we denoted
\begin{align}
B_{1\twobar}&\= 
\frac{\langle \Gamma_1,\Gamma_2 \rangle}{4}+
\langle \delta_1,\delta_2 \rangle 
\,, 
\qquad
B_{12}\=
\frac{\langle \Gamma_1,\Gamma_2 \rangle}{4}-
\langle \delta_1,\delta_2 \rangle  
\,, 
\qquad
A_{1\onebar} = 
\langle \Gamma_1 , \delta_1 \rangle ,
\\
A_{1 \twobar}&\= 
\frac{\langle \Gamma_1,\delta_2 \rangle-\langle \delta_1,\Gamma_2 \rangle}{2} 
\,, 
\qquad
A_{1 2}\= 
\frac{\langle \Gamma_1,\delta_2 \rangle+\langle \delta_1,\Gamma_2 \rangle}{2}
\,, 
\qquad 
A_{2\twobar} \= 
\langle \Gamma_2 , \delta_2 \rangle \,,
\\ 
c_1 &\= 
\frac{\beta}{4\pi} +  h_0  \delta_1^0  \,,
\qquad
c_2\= 
\frac{\beta}{4\pi} +  h_0  \delta_2^0   \,, 
\\
d_1  &\= 
\frac{-P^0 h_0 }{2} 
\,, \qquad 
d_2 \= 
\frac{P^0 h_0 }{2} 
\=  - d_1
\,. 
\label{eq:defd1d2}
\end{align}
\begin{figure}
    \centering
    \includegraphics[height=7.5cm]{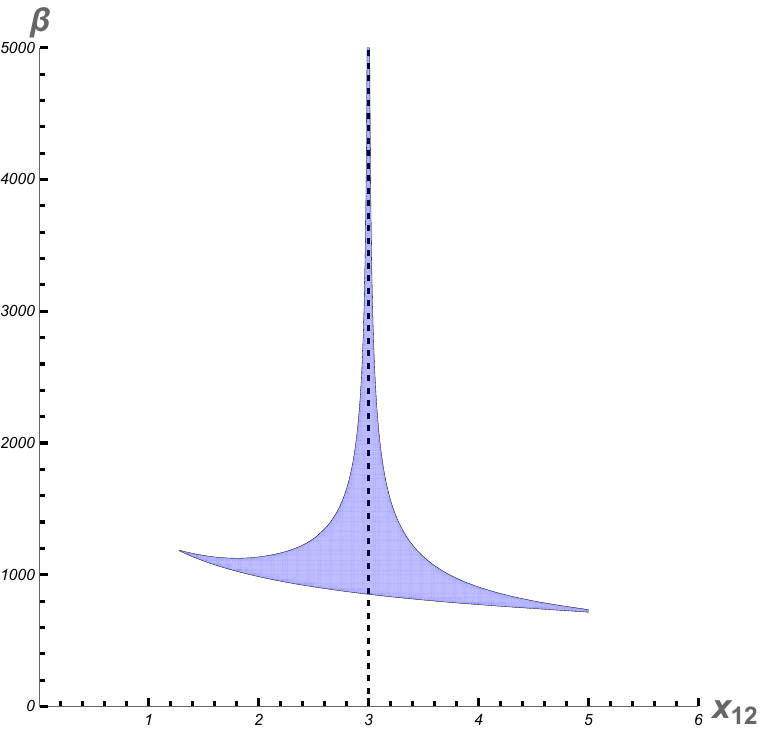}
    \caption{Example of allowable values of $x_{12}$ distance as a function of $\beta$ for the case of two BMPV black holes in AdS$_3 \times S^2$ as determined by smoothness of the saddle. The dashed lines represent the distance which matches the extremal bound state distance $x_{12}^{*} = -\langle \Gamma_1 , \Gamma_2 \rangle/\langle \Gamma_1, h \rangle$. 
     For the charges $\Gamma_1 = (-2,12,3,3) $, $\Gamma_2 = (2,6,3,3)$, as we lower the temperature, the space of solutions ``shrinks" until it collapses on the extremal value of the distance. For high temperatures, we pass the critical temperature after which the saddle cannot be smooth, and, presumably, we lose its contribution to the path integral.}
    \label{fig:TwoBMPVinAdS3}
\end{figure}
The allowable range of $x_{12}$ for which the above distances can be embedded in $\mathbb{R}^3$ is determined through the Cayley-Menger condition. 
To provide an example of a system in which a solution exists, 
we plot in Figure \ref{fig:TwoBMPVinAdS3} the region for which the Cayley-Menger determinant is positive for the charges $\Gamma_1 = (2,6,3,3)$, $\Gamma_2 = (-2,12,3,3) $.
Within this region, it is possible to then explicitly determine the Cartesian coordinates of the poles in $\mathbb{R}^3$ base space.

The saddle is now valid for temperatures for which all distances are positive, satisfy triangle inequalities, and have a positive Cayley-Menger determinant. When the saddle contributes, its on-shell action evaluates to the sum of black hole extremal entropies \cite{Boruch:2025biv}
\be 
- I_{\text{on-shell}}^{2\,\text{BMPV}}(\Gamma_\text{total}) = \pi \Sigma_{\text{total}} \= \pi \Sigma (\Gamma_1) + \pi \Sigma(\Gamma_2) \,,
\ee
with the explicit expressions determined from \eqref{eq:Q_function_Shmakova}. 

Note that choosing vanishing individual D6 charges ($P^0=0$) would result in $\langle \Gamma_1 , h\rangle =0$, 
which is precisely the wall-crossing condition. 
As has been shown in \cite{Boruch:2025biv}, at that point, the only configurations that can be embedded in $\mathbb{R}^3$ correspond 
to black holes infinitely far away from each other. 
This extends the known intuition at zero temperature that it is impossible to fit two BTZ black holes inside of AdS$_3 \times S^2$ \cite{deBoer:2008fk}.

\section{The moduli space of $5d$ index saddles}
\label{sec:moduli_space_discussion}

Above, we have described several saddle-point contributions to the supersymmetric index, each of which has its own moduli space of solutions. 
For example, in the case of the black rings and black lenses studied in Section~\ref{sec:index-saddles_flatspace} in asymptotically flat space, and the BMPV and black lens saddles studied in Section~\ref{sec:indices-in-ads3-s2} in asymptotically AdS$_3\times S^2$, the saddles are rigid after fixing translations and rotations. 
Thus, after accounting for rotations, the moduli space is three-dimensional. 
In the case of the black Saturn and black lens studied in Section~\ref{sec:index-saddles_flatspace}, 
and two BMPV black hole and black lens configurations studied in Section~\ref{sec:indices-in-ads3-s2}, the saddles have one undetermined parameter (parametrized by the base-space distance between the two north poles) after fixing translations and rotations. 
Thus, after accounting for rotations, the moduli space is a four-dimensional symplectic manifold~\cite{Boruch:2025biv}. 
Given these results, we can ask: What are the properties of the moduli space for generic saddles?

To answer this, we will follow the analysis of the moduli space for $4d$ black holes presented in \cite{Boruch:2025biv}. The moduli space is determined by the smoothness and integrability conditions, 
\begin{align}
\label{eq:smoothness}
\i \langle \gamma_i , H(\xvec_i) \rangle  \= \frac{\beta}{4\pi} \,, 
\qquad 
\i \langle \tilde \gamma_i , H(\tilde \xvec_i) \rangle \= -\frac{\beta}{4\pi} , \qquad & i \= 1, \dots, N_{BH}\,,
\\
\langle \Gamma_{\text{NUT}, i} , H(\xvec_{\text{NUT}, i}) \rangle \= 0, \qquad & i \= 1, \dots, N_{NUT} \,, 
\label{eq:integrability}
\end{align}
which have to be imposed at each of the two poles associated with each $4d$ black hole \eqref{eq:smoothness} and to each pole associated with a NUT charge \eqref{eq:integrability}. 
The number of linearly independent equations among \eqref{eq:smoothness} and \eqref{eq:integrability} are $4N_{BH} + 2N_{NUT} - 3 $. 
After fixing translations and before imposing the smoothness and integrability equations, the saddles are parametrized by  
$ 6N_{BH} + 3N_{NUT} - 3 $
real coordinates. In total, the moduli space dimension thus has 
\be 
N_\text{moduli} \= 2N_{BH} + N_{NUT} 
\ee
real dimensions. 
After also fixing rotations, this agrees with the dimension of the moduli space  
of the various cases studied 
in Sections~\ref{sec:index-saddles_flatspace} and \ref{sec:indices-in-ads3-s2}.  

We can also describe the boundary of the moduli space by finding points where four of the points among the black hole poles or the NUT poles become coplanar in basespace. 
The boundary itself has singular points at which more of the poles become coplanar. 
The most extreme case is when all the poles associated with the $4d$ black hole and NUT charges become coplanar.
After fixing translations and rotations, such configurations are parametrized by 
\be 
4N_{BH} + 2N_{NUT} - \underbrace{2}_{\text{Fixing CM}}- \underbrace{1}_{\substack{\text{Fixing rotations}\\ \text{ in the plane}}},
\nonumber
\ee
coordinates. This precisely agrees with the number of linearly independent equations among~\eqref{eq:smoothness} and~\eqref{eq:integrability};  thus, there is a discrete set of coplanar solutions that are isolated singular points on the boundary of the moduli space.

\section{Discussion}
\label{sec:discussion}

We have uncovered a variety of new gravitational saddles that contribute to the gravitational index in $5d$ supergravity, both in asymptotically flat space and in AdS$_3\times S^2$. 
For generic points in the moduli space of scalar boundary conditions and generic charge assignments, saddles with different horizon topologies ($S^3$, $S^1\times S^2$, or lens spaces $S^3/\mathbb{Z}_2$) can all contribute to the same index.
In some instances, single-centered black holes with spherical horizon topology provide the dominant contribution, 
while in others, black rings (with $S^1\times S^2$ horizons), black lenses (with lens-space horizons), 
or composite configurations involving several of these objects provide the dominant saddles. Compared to previous analyses, our results offer a systematic framework for evaluating and comparing the contributions of all such black objects to the index.
There are numerous lessons and open questions that arise from our analysis. 

\paragraph{The temperature dependence of the moduli space.} 
An intriguing feature of the five-dimensional index saddles is their distinct dependence on temperature. 
Although, as mentioned earlier, several saddles may contribute to the same supersymmetric index, their domains of existence are not identical: 
each configuration is supported only below a critical temperature determined by smoothness and the positivity of 
Cayley-Menger determinants in the underlying geometry. 
For example, the supersymmetric black ring persists only at sufficiently low temperatures, but as the temperature increases to a critical value~\eqref{eq:three-center_temperature_condition_2}, the triangle inequalities between the three poles of the harmonic function saturate, and the ring collapses onto the NUT charge. 
When this happens, the black ring saddle ceases to exist as a smooth configuration.\footnote{This assumes that the set of saddles that are picked up by the gravitational path integral should have a real embedding in base-space.} 
A similar phenomenon could be observed for the black hole and black lens. However, the temperatures at which these transitions occur are now different (see~\eqref{eq:BlackHole_cric_temperature} and~\eqref{eq:three-center_temperature_condition_2}, respectively). 
While this phenomenon occurs for the asymptotically flat space saddle that we analyzed, 
it does not occur for black holes in AdS$_3\times S^2$ when varying the AdS$_3$ temperature.\footnote{Recall that the temperature of the asymptotically flat space region of all these saddles scales to zero.} While this phenomenon occurs for the asymptotically flat space saddle that we analyzed, the case of BTZ black holes \cite{Boruch:2025qdq} suggests that it might not occur for black holes in AdS$_3\times S^2$ when varying the AdS$_3$ temperature. The fact that this phenomenon is observed in flat space and not in AdS, perhaps has to do with the existence of scattering states in the former that the helicity super-trace can be sensitive to. The contribution of such scattering states can make the helicity super-trace temperature dependent. A speculative interpretation of the disappearance of these saddles is therefore that, similar to the traditional wall-crossing phenomenon where the index changes discontinuously when varying the moduli, the helicity super-trace changes discontinuously at the critical temperature and receives a greater contribution from scattering states for temperatures greater than the critical temperature. 

\paragraph{Lessons about multi-center black hole saddles in AdS.} Our results hint at a broader lesson for AdS black holes. 
Through our calculation in AdS$_3 \times S^2$, we see that two BTZ black holes (with $S^1 \times S^2$ horizon topology) cannot form a smooth multi-center saddle, even when computing the index at finite temperature. 
However, when allowing a different fibration, we have found bound states of BMPV black holes (with $S^3$ horizon topology) and black lenses. In principle, all such contributions should be detected in the elliptic genus of the dual SCFT$_2$. This suggests that similar structures might arise in AdS$_d \times \mathcal M$ with $d>3$ when looking for solutions also supported on $\mathcal M$. For instance, it would be interesting to investigate whether ten-dimensional black holes with different horizon topologies can contribute to the type IIB supergravity index in asymptotically AdS$_5 \times S^5$, and if so, whether such contributions admit an interpretation in terms of protected states of $\mathcal{N}=4$ SYM. Our work also suggests an algorithm to search for such solutions: start from asymptotically $10d$ flat space saddles in type IIB supergravity that admit Killing spinors, then carefully take the decoupling limit such that the AdS$_5$ temperature remains finite to obtain the general asymptotically AdS$_5$ saddles that contribute to the index.

\paragraph{Finding more general $5d$ saddles.}  By  construction, all our solutions admit a $U(1)$ isometry generated by $\partial_\psi$, corresponding to translation along the M-theory circle. 
This feature is directly inherited from the $4d$/$5d$ lift, since the metric ansatz \eqref{eq:5d_uplift_formula_1} explicitly has $\partial_\psi$ as a Killing vector. 
However, this symmetry may represent a significant restriction: by imposing that $\partial_\psi$ is a Killing vector, we could be excluding more general families of supersymmetric configurations that do not arise as direct uplifts of four-dimensional index geometries. 
Such saddles would go beyond the standard multi-center ansatz and could potentially capture contributions to the index not visible within the current uplift framework. 
More generally, one may follow the approach of \cite{Tod:1983pm, Gauntlett:2002nw} for characterizing the full class of asymptotically flat spacetimes that admit Killing spinors in $5d$.

\subsection*{Acknowledgments}
We especially thank Gustavo J. Turiaci for valuable discussions during the early stages of the project.  
We are also grateful to Davide Cassani, Spencer Tamagni, Enrico Turetta, and Alejandro Ruipérez Vicente for useful discussions. RE was
supported by the grants MICINN PID2022-136224NB-C22, AGAUR 2021 SGR 00872, and CEX2024-001451-M from MICIU/AEI 10.13039/501100011033. LVI and JB was supported by the DOE Early Career Award DE-SC0025522 and by the DOE Grant DE-SC0019380.

\bibliographystyle{utphys2}
{\small \bibliography{Biblio}{}}

\end{document}